\documentclass[preprint,12pt,authoryear]{elsarticle}
\usepackage{textcomp}
\usepackage{amssymb,amsfonts,amsthm}
\usepackage{geometry}
\usepackage{subfigure}
\usepackage{hyperref}
\usepackage[fleqn]{amsmath}
\usepackage{natbib}

\usepackage{booktabs}
\usepackage{gensymb}
\usepackage{graphicx}
\usepackage{color}
\usepackage{xcolor}
\usepackage{lineno}
\usepackage{bm}

\def \div{\mbox{div\hskip 1pt}}
\def \Div{\mbox{Div\hskip 1pt}}
\def \tr{\mbox{tr\hskip 1pt}}
\def \grad{\mbox{grad\hskip 1pt}}
\def \Grad{\mbox{Grad\hskip 1pt}}

\begin{document}
\numberwithin{equation}{section}
\begin{frontmatter}
\title{Buckling of residually stressed cylindrical tubes under compression}

\author[add1]{Tao Zhang}
\author[add2]{Luis Dorfmann}
\author[add3,add1]{Yang Liu\corref{cor1}}
\ead{liuy3@maths.ox.ac.uk, tracy\_liu@tju.edu.cn} 
\address[add1]{Department of Mechanics, School of Mechanical Engineering, Tianjin University, Tianjin 300350, China}
\address[add2]{Department of Civil and Environmental Engineering, Tufts University, Medford, MA, USA}
\address[add3]{Mathematical Institute, University of Oxford, Oxford, OX2 6GG, UK}
%\address[add4]{National Key Laboratory of Vehicle Power System, Tianjin 300350, China}

\cortext[cor1]{Corresponding author.}

\begin{abstract}

We evaluate the loss of stability  of axially compressed slender and thick-walled tubes subject to a residual stress distribution. The nonlinear theory of elasticity, when used to analyze the underlying deformation, shows  that the residual stress induces  preferred directions in the reference configuration. The incremental theory, given in  Stroh form, is used to derive an exact bifurcation condition. The critical stretch and the associated critical buckling mode are identified for  axisymmetric and asymmetric increments in the deformation. Mode transitions are illustrated as the tube slenderness varies. For slender tubes, Euler buckling is energetically favorable, and the effect of residual stress is  negligible. However, for short and thick-walled tubes where barreling mode is dominant, the residual stress significantly affects the buckling behavior and may eliminate barreling instability. We show that, depending on its magnitude and direction, residual stress can either accelerate or delay instability. Phase diagrams for various modes are obtained and provide insight into pattern selection across different tube geometries. 
\end{abstract}

%\maketitle
\end{frontmatter}

%%%%%%%%%%%%%%%%%%%%%%%%%%%%%%%%%%%%%%%%%%%%%%%%%%%%%%%%%%%%%%%
\section{Introduction}
%%%%%%%%%%%%%%%%%%%%%%%%%%%%%%%%%%%%%%%%%%%%%%%%%%%%%%%%%%%%%%% 

Tubular structures are frequently encountered in both biology  and engineering applications—for example, blood vessels \citep{fung1993mechanical,SOKOLIS2015229}, microtubules in cells \citep{brangwynne2006microtubules,li2008mechanics}, packers used in petroleum engineering \citep{liu2018axial,lan2019sealing}, and cellular foams \citep{chawla2022superior}, among others. In practice, these structures may undergo finite deformations, making them susceptible to buckling under combined axial and transverse loads. In a seminal study, \citet{koiter1945stability} developed a post-buckling theory and demonstrated that imperfection sensitivity can cause an axially compressed cylindrical shell to collapse earlier than theoretical prediction. Since then, the stability and bifurcation behavior of cylindrical tubes or shells under various load conditions have been a topic of fundamental interest in both the mechanics and engineering communities.

For thick and incompressible neo-Hookean tubes subjected to end loads, \citet{Wilkes1955AM} employed finite deformation theory to investigate buckling instability. An important finding was that the associated eigenvalue problem arising from a linear bifurcation analysis in cylindrical polar coordinates can be solved exactly in terms of Bessel functions. Decades later, this classical problem was revisited by \citet{pan1997instabilitya, pan1997remarks, pan1999instabilityb} and results for different material models were provided. Considering material compressibility, \citet{DorfmannA2006} explored the full bifurcation landscape for  tubes under axial compression and unveiled two distinct forms of the barreling mode. \citet{goriely2008nonlinear} re-derived the exact bifurcation condition for incompressible tubes of arbitrary wall thickness and length under compression using the Stroh formalism. They also obtained useful asymptotic solutions and constructed a phase diagram characterizing different bifurcation modes. \citet{springhetti2023buckling} re-established the classical  conditions for thin-walled cylinders and proposed a generalized framework applicable to a broad class of nonlinear hyperelastic materials. 

These studies focused on using the nonlinear theory to determine the critical buckling loads and corresponding  mode shapes. However, they do not provide insight into the post-buckling evolution or the nature of the bifurcation. To address this gap, \citet{dai2015pitchfork} adopted a coupled series–asymptotic expansion  to derive a reduced model applicable for thin-walled tubes. Focusing on asymmetric bifurcation modes, they obtained analytical post-buckling solutions and revealed the counterintuitive result that a thick-walled tube may be softer than a thin one. Recently, \citet{zhou2023three}  performed a post-buckling analysis of axially compressed thick-walled tubes in a fully three-dimensional setting to highlight the influence of geometric parameters on both buckling and post-buckling behaviors.

In many practical applications, tubular structures are subjected to complex loading conditions.  To advance the understanding of tube stability under such conditions, \citet{haughton1979bifurcationa,haughton1979bifurcationb} developed an analysis for both thin- and thick-walled tubes subjected to axial force and internal pressure. They examined prismatic, axisymmetric, and asymmetric buckling modes, and showed how geometric parameters, such as wall thickness, length-to-radius ratio, and axial stretch influence the onset of instability. \citet{zhu2008asymmetric} provided a comprehensive bifurcation analysis of thick-walled tubes under combined axial load and external pressure, capturing a rich spectrum of asymmetric and axisymmetric modes across a wide range of wall thicknesses. \citet{chen2017bifurcation} used the incremental theory to investigate instabilities of functionally graded tubes under combined axial stretch and internal or external pressures. Their parametric study highlights how graded material properties and loading conditions affect bifurcation behavior, enabling the design of soft cylindrical structures with tailored mechanical responses.  \citet{liu2018axial} explored  buckling of hyperelastic tubes under constrained compression and identified transition between axial and circumferential instabilities. 

In addition to stress generated by external loads, a residual stress  may develop in biological tissue due to growth and remodeling and  in synthetic materials due to discontinuities between different components and the fabrication process. These additional stresses are in equilibrium with zero surface traction and persist even after external loads are removed. A well-known example is arterial tissue, which exhibits axial, circumferential, and radial residual stress components  \citep{vaishnav1987residual, holzapfel2010modelling}. It turns out that residual stress in arteries serves to homogenize stress distributions induced by blood pressure \citep{liu1989relationship, fung1991residual,fung1993mechanical}. Studies including the effect of residual stress in cardiac tissue  have been carried  in \cite{omens1990residual, costa1997three,rausch2013effect}

It is well known that  the mechanical  properties of residually stressed biological tissue are deformation dependent. To capture this behavior, \citet{hoger1985residual, hoger1993constitutive, hoger1996elasticity, hoger1997virtual} developed a nonlinear elasticity theory to account for finite deformation and  residual stress. Using Hoger's work as starting point, \citet{merodio2013influence} analyzed the influence of residual stress on the finite deformation  of hyperelastic materials, focusing on problems such as simple shear, tube inflation, and tension–torsion of cylinders. \citet{ahamed2016modelling} investigated how residual stress affects wall stress distribution in abdominal aortic aneurysms (AAAs), using patient-specific data and modeling the material as fiber-reinforced. \citet{dorfmann2021effect} studied the impact of residual stress on the stability of pressurized tubes under axial extension, focusing on axisymmetric modes. \citet{melnikov2021bifurcation} conducted a detailed bifurcation analysis of residually stressed tubes subjected to combined extension and inflation, considering both axisymmetric and asymmetric responses. More recently, \citet{liu2024localized} demonstrated that residual stress can trigger localized instabilities in a stretched cylinder and derived an analytical bifurcation condition for localization. Furthermore, both localized bulging and necking may occur, depending on the loading strategy \citep{liu2024localized,liu2024reduced}.

The focus of this work is provide a comprehensive investigation of the bifurcation behavior of  tubes under axial compression, with particular attention to both axial and circumferential instability modes. We specialize the incremental equations and boundary conditions in  Stroh form to derive an exact bifurcation condition and to investigate  how residual stress influences the onset of buckling. The rest of this paper is organized as follows. Section \ref{fundamental equations} summarizes the fundamental theory for residually stressed solids. Section \ref{Incremental theory} outlines the incremental theory used for the bifurcation analysis. In Section \ref{Application to a compressed cylindrical tube}, we examine  compression of a residually stressed cylindrical tube,  characterize the primary deformation, and derive the bifurcation condition using the Stroh formalism and surface impedance matrix method. A comprehensive parametric analysis of various buckling modes is presented in Section \ref{Parametric study}. Concluding remarks are given in Section \ref{conclusion}.

%%%%%%%%%%%%%%%%%%%%%%%%%%%%%%%%%%%%%%%%%%%%%%%%%%%%%%%%%%%%%%%
\section{Fundamental theory}\label{fundamental equations}
%%%%%%%%%%%%%%%%%%%%%%%%%%%%%%%%%%%%%%%%%%%%%%%%%%%%%%%%%%%%%%%

The objective of this work is to investigate  buckling instability of a compressed hyperelastic tube in the presence of residual stress. It is shown by  \cite{rajagopal2024residual,bustamante2024residual} that residual stress or prestress in an elastic body can significantly alter its effective material symmetry, leading to an anisotropic mechanical response, even when the material is initially isotropic.

We consider a residually stressed body which, without external mechanical loads, occupies the reference configuration $\mathcal{B}_{\mathrm r}$. When subjected to mechanical loads, the body undergoes finite deformation to occupy the current configuration,  denoted $\mathcal{B}_{\mathrm e}$. We introduce the position vectors $\mathbf{X}$ and $\mathbf{x}$ to identify a material point in the reference and current configurations,  respectively. The deformation from $\mathcal{B}_{\mathrm r}$ to $\mathcal{B}_{\mathrm e}$ is written as $\mathbf{x}=\boldsymbol\chi(\mathbf{X})$. The deformation gradient $\mathbf{F}$ is given by $\mathbf{F}=\Grad\boldsymbol\chi$, where $\Grad$ is the gradient operator with respect to $\mathbf{X}$. The associated right and left Cauchy-Green deformation tensors, denoted  $\mathbf{C}$ and $\mathbf{B}$, respectively, are defined by
\begin{equation}
	\mathbf{C}=\mathbf{F}^{\mathrm{T}}\mathbf{F}\ \ \ \ \mathbf{B}=\mathbf{F}\mathbf{F}^{\mathrm{T}}, \label{eq:right-left Cauchy-green tensor}
\end{equation}
where the superscript $^\mathrm{T}$ indicates  the transpose of a second-order tensor. For an incompressible 
material, we have
\begin{equation}
	J=\det\mathbf F =1.\label{eq:J}
\end{equation}

%%%%%%%%%%%%%%%%%%%%%%%%%%%%%%%%%%%%%%%%%%%%%%%%%%%%%%%%%%%%%%%
\subsection{Equilibrium and residual stress}
%%%%%%%%%%%%%%%%%%%%%%%%%%%%%%%%%%%%%%%%%%%%%%%%%%%%%%%%%%%%%%%

 We assume that there are no body forces  and no intrinsic couple stresses. Then, the Cauchy stress $\boldsymbol \sigma$ is symmetric and  satisfies the equilibrium equation
\begin{equation}
	\div\boldsymbol{\sigma}=\mathbf{0},\label{eq:equilibrium equation}
\end{equation}
where $\div$ is the divergence operator with respect to $\mathbf{x}\in\mathcal{B}_{\mathrm e}$. The mechanical traction  on the current boundary $\partial\mathcal{B}_{\mathrm e}$ is specified by
\begin{equation}                                    
\boldsymbol{\sigma}\mathbf{n}=\mathbf{t}_{\mathrm a}\quad\mathrm{on}~\partial\mathcal{B}_{\mathrm e},\label{eq:boundary condition}
\end{equation}
where $\mathbf{n}$ denotes the outward unit normal vector.

To obtain the equivalent Lagrangian form, we first recall  Nanson's formula \citep{ogden1997non} 
\begin{equation}
	\mathbf{n}\mathrm{d}s=J \mathbf{F}^{-\mathrm T} \mathbf{N}\mathrm{d}S,\label{eq:nanson formula}
\end{equation}
which connects an infinitesimal area element $\mathrm{d}s$ with outward unit normal vector $\mathbf{n}$ in the current configuration to the corresponding area element $\mathrm{d}S$ with outward unit normal vector $\mathbf{N}$ in the 
reference configuration. Using equations \eqref{eq:boundary condition} and \eqref{eq:nanson formula}, we find that
\begin{equation}
	\mathbf t_{\mathrm a}\mathrm{d}s=\boldsymbol{\sigma}\mathbf{n}\mathrm{d}s=J\mathbf{F}^{-1}\boldsymbol{\sigma}\mathbf{N}\mathrm{d}S=\mathbf{T}^\mathrm{T}\mathbf{N}\mathrm{d}S,\label{eq:boundary condition(rc)}
\end{equation}
where $\mathbf T$ denotes the nominal stress
\begin{equation}
	\mathbf{T}=J\mathbf{F}^{-1}\boldsymbol{\sigma}.\label{eq:nominal stress}
\end{equation}

It  satisfies the equilibrium equation
\begin{equation}
	\Div \mathbf{T}=\mathbf{0},\label{eq:equilibrium eq(nominal)}
\end{equation}
where $\Div$ denotes the divergence operator defined in the reference configuration $\mathcal{B}_{\mathrm r}$. The Lagrangian form of the traction boundary condition reads
\begin{equation}
	\mathbf{T}^\mathrm T \mathbf{N}=\mathbf{t}_{\mathrm A},\quad\mathrm{on}~  \partial \mathcal{B}_{\mathrm r},\label{eq:traction bc}
\end{equation}
where $\mathbf{t}_{\mathrm A}$ is the load per unit surface area on $\mathcal B_{\mathrm r}$ and $\mathbf{N}$ is the outward unit normal vector.

Let the reference configuration $\mathcal{B}_{\mathrm r}$ have a residual stress   $\boldsymbol{\tau}$. Then, there is no distinction
between different measures and we have  $\mathbf{T}=\boldsymbol{\sigma}=\boldsymbol{\tau} $. From \eqref{eq:equilibrium eq(nominal)} we obtain
\begin{equation}
	\Div \boldsymbol{\tau}=\mathbf 0, \label{eq:equilibrium equation(residual)}
\end{equation}
with zero traction forces. Hence \eqref{eq:traction bc} gives
\begin{equation}
	\boldsymbol{\tau}\mathbf{N}=\mathbf{0},\ \ \ \mathrm{on} ~ \partial \mathcal{B}_{\mathrm r},\label{eq:residual bc}
\end{equation}
with $\boldsymbol{\tau}=\boldsymbol{\tau}^{\mathrm T}$.

%%%%%%%%%%%%%%%%%%%%%%%%%%%%%%%%%%%%%%%%%%%%%%%%%%%%%%%%%%%%%%%
\subsection{Constitutive equations}
%%%%%%%%%%%%%%%%%%%%%%%%%%%%%%%%%%%%%%%%%%%%%%%%%%%%%%%%%%%%%%%

For a residually stressed elastic solid, the strain energy density per unit volume is a function of the deformation 
gradient $\mathbf{F}$ and the residual stress $\boldsymbol{\tau}$, which is written as $W(\mathbf{F},\boldsymbol{\tau})$. Note that $W$ is objective since it depends on $\mathbf{F}$ only via the right Cauchy-Green deformation tensor $\mathbf{C}$ and $\boldsymbol{\tau}$ is unaffected by rotations in the deformed configuration $\mathcal{B}_{\mathrm e}$. The Cauchy  and the nominal stresses  are given  by
\begin{equation}
	\boldsymbol\sigma=\mathbf{F}\frac{\partial W}{\partial\mathbf{F}}(\mathbf{F},\boldsymbol{\tau})-p\mathbf{I},\quad\mathbf{T}=\frac{\partial W}{\partial\mathbf{F}}(\mathbf{F},\boldsymbol{\tau})-p\mathbf{F}^{-1},\label{eq:Cauchy stress and nominal stress}
\end{equation}
respectively, where $p$ is the Lagrange multiplier associated with the incompressibility constraint \eqref{eq:J}. In $\mathcal{B}_{\mathrm r}$, 
  when $\mathbf{F}$ = $\mathbf{I}$, equation \eqref{eq:Cauchy stress and nominal stress} reduces to
  \begin{equation}
	\boldsymbol{\tau}=\frac{\partial W}{\partial\mathbf{F}}(\mathbf{I},\boldsymbol{\tau})-p_{\mathrm r}\mathbf{I},\label{eq:tau}
  \end{equation}
where $p_{\mathrm r}$ is the Lagrange multiplier in $\mathcal{B}_{\mathrm r}$.

According to the continuum theory of anisotropic materials given by \citet{spencer1971theory} and generalized to residually stressed materials \citep{hoger1996elasticity,hoger1997virtual}, the strain energy function $W(\mathbf{C},\boldsymbol{\tau})$ depends on $9$ independent invariants of $\mathbf{C}$, $\boldsymbol{\tau}$ and their combinations.
The principal invariants of $\mathbf{C}$ are given by
 \begin{equation}
    I_1=\operatorname{tr} \mathbf{C},\ \ \ I_2=\frac{1}{2}\left[I_{1}^{2}-\operatorname{tr} \left(\mathbf{C}^{2}\right)\right],\ \ \ I_{3}=\det\mathbf{C},\label{eq:principal invariants for c}
 \end{equation}
and $I_{3}=1$ according to the incompressibility condition \eqref{eq:J}. The  invariants for $\boldsymbol{\tau}$ independent of the deformation are collectively denoted by
\begin{equation}
    I_{4}=\{\operatorname{tr} \boldsymbol{\tau},\ \frac{1}{2}\left[(\operatorname{tr} \boldsymbol{\tau})^{2}-\operatorname{tr} \left(\boldsymbol{\tau}^{2}\right)\right],\ \det \boldsymbol{\tau}\}.\label{eq:principal invariants for tau}
\end{equation}
There exist four additional invariants that depend on $\mathbf{C}$ and $\boldsymbol{\tau}$  defined by
\begin{equation}
    I_5=\operatorname{tr}\left(\boldsymbol{\tau}\mathbf{C}\right),\quad I_6=\operatorname{tr}\left(\boldsymbol{\tau}\mathbf{C}^2\right),\quad I_7=\operatorname{tr}\left(\boldsymbol{\tau}^2\mathbf{C}\right),\quad I_8=\operatorname{tr}\left(\boldsymbol{\tau}^2\mathbf{C}^2\right).\label{eq:principal invariants for tau and c}
\end{equation}

To indicate the dependence of $W$ on the invariants  \eqref{eq:principal invariants for c}-\eqref{eq:principal invariants for tau and c} we write
\begin{equation}
	W=W(I_1,I_2,I_4,I_5,I_6,I_7,I_8),\label{eq:W1}
\end{equation}
where $I_3=1$ and is  therefore excluded. The nominal and Cauchy stresses are the obtained by
\begin{equation}
	\mathbf{T}=\sum_{i\in\mathcal{I}}W_{i}\frac{\partial I_{i}}{\partial\mathbf{F}}-p\mathbf{F}^{-1},\ \ \ \boldsymbol{\sigma}=\mathbf{F}\mathbf{T},
\end{equation}
where $W_{i}=\partial W/\partial I_{i}$ and $\mathcal{I}=\{1,2,5,6,7,8\}$. Expanding gives the explicit expression of the Cauchy
stress tensor 
\begin{align}
    \notag \boldsymbol{\sigma}=&~2W_{1}\mathbf{B}+2W_{2}(I_{1}\mathbf{B}-\mathbf{B}^{2})+2W_{5}\boldsymbol{\Sigma}+2W_{6}(\boldsymbol{\Sigma}\mathbf{B}+\mathbf{B}\boldsymbol{\Sigma})\\&+2W_{7}\boldsymbol\Xi+2W_{8}(\boldsymbol\Xi\mathbf{B}+\mathbf{B}\boldsymbol\Xi)-p\mathbf{I}.\label{eq:Cauchy stress tensor}
\end{align}
where we introduced the short-hand notations $\boldsymbol\Sigma=\mathbf{F}\boldsymbol{\tau}\mathbf{F}^{\mathrm{T}}$ and $\boldsymbol\Xi=\mathbf{F}\boldsymbol{\tau}^{2}\mathbf{F}^{\mathrm{T}}$.

In the reference configuration $\mathcal{B}_{\mathrm r}$, the invariants in \eqref{eq:principal invariants for c} and \eqref{eq:principal invariants for tau and c} assume the values
\begin{equation}
	I_1=I_2=3,\quad I_5=I_6=\operatorname{tr}\boldsymbol{\tau},\quad I_7=I_8=\operatorname{tr}\left(\boldsymbol{\tau}^2\right),\label{eq:in11}
\end{equation}
and  \eqref{eq:Cauchy stress tensor}  reduces to
\begin{equation}
	\boldsymbol{\tau}=(2W_1+4W_2-p_r)\mathbf{I}+2(W_5+2W_6)\boldsymbol{\tau}+2(W_7+2W_8)\boldsymbol{\tau}^2.\label{eq:tau special}
\end{equation}
This requires that  the energy function in $\mathcal{B}_{\mathrm r}$  is restricted to
\begin{equation}
	2W_1+4W_2-p_{\mathrm r}=0,\quad W_5+2W_6=\frac{1}{2},\quad W_7+2W_8=0,\label{eq:vanish}
\end{equation}
see, for example,  \cite{shams2011initial,dorfmann2021effect} for details.

%%%%%%%%%%%%%%%%%%%%%%%%%%%%%%%%%%%%%%%%%%%%%%%%%%
\section{Incremental theory}\label{Incremental theory}
%%%%%%%%%%%%%%%%%%%%%%%%%%%%%%%%%%%%%%%%%%%%%%%%%%

Here we provide a brief summery of the incremental theory and refer to, for example,  \cite{ogden1997non} for details. The buckling analysis is then reduced to studying the linearized increment in the deformation superimposed on a finitely deformed configuration subjected to a residual stress. 

Consider an increment in the displacement, denoted $\dot{\mathbf x} = \dot{\boldsymbol \chi}(\mathbf X)$, superimposed on the current configuration $\mathcal B_{\mathrm e}$ and resulting in an  increment in the deformation gradient $\dot{\mathbf F} = \Grad \dot{\boldsymbol{\chi}} $. We use a superposed dot to indicate an incremental quantity, see, for example,  \citet{haughton1979bifurcationb}, \citet{hoger1997virtual}, \citet{melnikov2021bifurcation}, \citet{dorfmann2021effect}, \citet{liu2024localized}.

For what follows, it is convenient to write the increment of the displacement $\dot{\mathbf x}$ in Eulerian form resulting in an increment as a function of $\mathbf x$. This is obtained via the transformation
\begin{equation}
	\mathbf{u}(\mathbf{x})=\mathbf{u}\left(\boldsymbol\chi(\mathbf{X})\right)=\dot{\mathbf{x}}(\mathbf{X}).
\end{equation}

The increment of equation \eqref{eq:J} gives $\dot{J} = J \, \text{tr}\left(\mathbf F^{-1} \dot{\mathbf F}\right)=0$, and we  obtain the  incremental form of the incompressibility condition
\begin{equation}
	\tr \mathbf{L} \equiv \div  \mathbf{u} = 0,\label{eq:imcompressible condition2}
\end{equation}
with
\begin{equation}
\label{gradu}
	\mathbf{L} = \dot{\mathbf{F}} \mathbf{F}^{-1} = \grad \mathbf{u}
\end{equation}
being the incremental displacement gradient, and $\grad$  the gradient operator with respect to $ \mathcal{B}_{\mathrm e} $.

The incremental deformation generates an increment in the nominal stress, denoted by $ \dot{\mathbf{T}} $, which satisfies the equation of equilibrium
\begin{equation}
	\Div  \dot{\mathbf{T}} = \mathbf{0},\label{eq:incremental equilibrium equation}
\end{equation}
and the associated incremental form of the boundary condition 
\begin{equation}	           
\dot{\mathbf{T}}^\mathrm{T}\mathbf{N}=\dot{\mathbf{t}}_{\mathrm A},\quad\mathrm{on~} \partial \mathcal{B}_{\mathrm r}.\label{eq:incremental bc}
\end{equation}\par
To write the incremental forms of the equilibrium equation and boundary conditions as a function of $\mathbf x$, we use the push-forward operation applied to $\dot{\mathbf{T}}$ and obtain $\dot{\mathbf{T}}_0 = \mathbf{F} \dot{\mathbf{T}}$. Accordingly,  \eqref{eq:incremental equilibrium equation} transforms to
\begin{equation}
\operatorname{div}\dot{\mathbf{T}}_0=\mathbf{0},\label{eq:incremental equilibrium}
\end{equation}
and  \eqref{eq:incremental bc} becomes
\begin{equation}
\dot{\mathbf{T}}_0^\mathrm{T}\mathbf{n}=\dot{\mathbf{t}}_{\mathrm a0},\quad\mathrm{on~\partial \mathcal{B}_{\mathrm e}},\label{eq:bc form}
\end{equation}
where $\dot{\mathbf{t}}_{\mathrm a0}$ represents the increment of the mechanical traction in the current configuration.

Let \{$\mathbf{e}_1$, $\mathbf{e}_2$, $\mathbf{e}_3$\} be an orthonormal basis for an orthogonal curvilinear coordinate system. Then, in component form, equation \eqref{eq:incremental equilibrium} yields the three scalar equations:
\begin{equation}
\dot{T}_{0ji,j} + \dot{T}_{0ji} \mathbf{e}_k \cdot \mathbf{e}_{j,k} + \dot{T}_{0kj} \mathbf{e}_i \cdot \mathbf{e}_{j,k} = 0, \quad i = 1, 2, 3,\label{eq:expand increment equilibrium equation}
\end{equation}
where subscripts following a comma indicate partial derivatives and where the standard summation convention over repeated indices applies. 

An increment in the deformation $\dot{\mathbf x}$ results in an incremental in the nominal stress. For an incompressible material this is given by
\begin{equation}	\dot{\mathbf{T}}=\mathcal{A}\dot{\mathbf{F}}+p\mathbf{F}^{-1}\dot{\mathbf{F}}\mathbf{F}^{-1}-\dot{p}\mathbf{F}^{-1},\label{eq:incremental nominal stress}
\end{equation}
where $\dot{p}$ is the increment of the Lagrange multiplier and $\mathcal{A}$  is the fourth-order elastic moduli tensor with components 
\begin{equation}
\mathcal{A}_{\alpha i\beta j}=\frac{\partial^2W}{\partial F_{i\alpha}\partial F_{j\beta}},\label{eq:elastic moduli tensor}
\end{equation}
with the symmetry $\mathcal{A}_{\alpha i\beta j}=\mathcal{A}_{\beta j\alpha i}$. The updated version of \eqref{eq:incremental nominal stress} reads
\begin{equation}
\dot{\mathbf{T}}_0=\boldsymbol{\mathcal{A}}_0\mathbf{L}+p\mathbf{L}-\dot{p}\mathbf{I},\label{eq:increment version}
\end{equation}
where the components of the updated elastic moduli tensor $\boldsymbol{\mathcal{A}}_0$ are calculated as
\begin{equation}
    \mathcal{A}_{0lkji}=\mathcal{A}_{0jilk}=F_{j\alpha}F_{l\beta}\mathcal{A}_{\alpha i\beta k}.
\end{equation}

In terms of invariants the updated elasticity tensor has the expanded component form
\begin{equation}
	\mathcal{A}_{0piqj}=\sum_{r\in\mathcal{I}}W_{r}F_{p\alpha}F_{q\beta}\frac{\partial^{2}I_{r}}{\partial F_{i\alpha}\partial F_{j\beta}}+\sum_{r,s\in\mathcal{I}}W_{rs}F_{p\alpha}F_{q\beta}\frac{\partial I_{r}}{\partial F_{i\alpha}}\frac{\partial I_{s}}{\partial F_{j\beta}},
\end{equation}
where $W_{ij}=\partial W/\partial I_i\partial I_j$ with $i,j\in\mathcal{I}$. The  expansion of $\mathcal{A}_{0piqj}$ is given in \citet{melnikov2021bifurcation}. We note that the components $\mathcal{A}_{0piqj}$ are functions of residual stress \citep{destrade2013stress}.

%%%%%%%%%%%%%%%%%%%%%%%%%%%%%%%%%%%%%%%%%%%%%%%%%%%%%%%
\section{Illustration}\label{Application to a compressed cylindrical tube}
%%%%%%%%%%%%%%%%%%%%%%%%%%%%%%%%%%%%%%%%%%%%%%%%%%%%%%%

To obtain explicit results, we  consider a residually stressed neo-Hookean material with the mechanical properties  specified by the energy function
\begin{equation}
	W=\frac12\mu(I_1-3)+\frac12(I_5-\operatorname{tr}\boldsymbol\tau)+\frac{1}{4}\xi(I_5-\operatorname{tr}\boldsymbol\tau)^2,
    \label{eq:nH}
\end{equation}
where $\mu>0$ is the shear modulus in the reference configuration and $\xi$ is a non-negative material constant. It is easy to see that \eqref{eq:nH} satisfies the restriction  in \eqref{eq:vanish}. Note that this model has been used by  \cite{dorfmann2021effect} to report on the restrictions imposed by the strong ellipticity condition and by \cite{liu2024localized} to analyze  localized instabilities of a residually stressed solid cylinder.

%%%%%%%%%%%%%%%%%%%%%%%%%%%%%%%%%%%%%%%%%%%%
\subsection{Primary deformation}
%%%%%%%%%%%%%%%%%%%%%%%%%%%%%%%%%%%%%%%%%%%%

Consider a homogeneous, hyperelastic and incompressible thick-walled tube whose reference configuration $\mathcal B_{\mathrm r}$ is given in terms of cylindrical polar coordinates  $(R,\Theta,Z)$ by 
\begin{equation}
A\leqslant R\leqslant B,\quad0\leqslant\Theta\leqslant2\pi,\quad0\leqslant Z\leqslant L,
\end{equation}
where  $A, B$ are  the inner and outer radii and $L$ is the reference length. The application of external loads results in the deformed configuration specified by  
\begin{equation}
a\leqslant r\leqslant b,\quad0\leqslant\theta\leqslant2\pi,\quad0\leqslant z\leqslant l,
\end{equation}
where   $(r,\theta,z)$ are the cylindrical polar coordinates, $a,b$  the inner and outer deformed radii, $l=\lambda_z L$ is the current length and $\lambda_z$ the axial stretch of the tube.  The corresponding deformation gradient  has the diagonal form
\begin{equation}
\mathbf{F}=\frac{\mathrm{d}r}{\mathrm{d}R}\mathbf{e}_{r}\otimes\mathbf{E}_{r}+\frac{r}{R}\mathbf{e}_{\theta}\otimes\mathbf{E}_{\theta}+\lambda_{z}\mathbf{e}_{z}\otimes\mathbf{E}_{z},
\end{equation}
where  $\{\mathbf{E}_r,\mathbf{E}_\theta,\mathbf{E}_z\}$ and $\{\mathbf{e}_r,\mathbf{e}_\theta,\mathbf{e}_z\}$  are the orthonormal basis vectors in the reference and current configurations, respectively. Using \eqref{eq:right-left Cauchy-green tensor}$_1$  gives the right Cauchy-Green deformation tensor 
\begin{equation}
\mathbf{C}=\frac{R^2}{r^2\lambda_z^2}\mathbf{E}_R\otimes\mathbf{E}_R+\frac{r^2}{R^2}\mathbf{E}_\Theta\otimes\mathbf{E}_\Theta+\lambda_z^2\ \mathbf{E}_Z\otimes\mathbf{E}_Z,
\label{eq:right-CG}
\end{equation}
where we used  the  incompressibility condition \eqref{eq:J} in the form
\begin{equation}
r\lambda_{z}\mathrm{d}r=R\mathrm{d}R.
\end{equation}
 This, after integration gives 
\begin{equation}
    r^2=\lambda_{z}^{-1}(R^2-A^2)+a^2,\label{eq:r with R}
\end{equation}
where $a$ is  to be determined. 

Consider an axisymmetric residual stress distribution in $\mathcal{B}_{\mathrm r}$ with nonzero components $\tau_{RR}$ and $\tau_{\Theta\Theta}$.  Then,  \eqref{eq:equilibrium equation(residual)}  reduces to the non-zero component equation
\begin{equation}
\frac{\mathrm{d}\tau_{RR}}{\mathrm{d} R}+\frac1R(\tau_{RR}-\tau_{\Theta\Theta})=0,\label{eq:43}
\end{equation}
and the zero traction boundary condition \eqref{eq:residual bc} specializes to
\begin{equation}
\tau_{RR}=0,\quad\mathrm{on} ~\partial \mathcal B_{\mathrm r}\label{eq:44}
\end{equation}

The invariants $I_1,I_5$ used  in the energy function \eqref{eq:nH} have the forms 
\begin{equation}
	\begin{aligned}
		I_1 &= \frac{R^2}{r^2\lambda_z^2} +\frac{r^2}{R^2}+\lambda_z^2, \\
		I_5 &=\frac{R^2}{r^2 \lambda_z^2}\tau_{RR}+\frac{r^2}{R^2}\tau_{\Theta\Theta},
	\end{aligned}
\end{equation}
which are used to evaluate the components of the  Cauchy stress \eqref{eq:Cauchy stress tensor} 
\begin{equation}
	\begin{aligned}
        &\sigma_{rr} =\mu \frac{R^2}{r^2\lambda_z^2}+\left\{1+\xi \left[\left(\frac{R^2}{r^2 \lambda_z^2}-1\right)\tau_{RR}+\left(\frac{r^2}{R^2}-1\right)\tau_{\Theta\Theta}\right]\right\}\frac{R^2}{r^2\lambda_z^2}\tau_{RR}-p, \\
		&\sigma_{\theta\theta} = \mu \frac{r^2}{R^2}+\left\{1+\xi \left[\left(\frac{R^2}{r^2 \lambda_z^2}-1\right)\tau_{RR}+\left(\frac{r^2}{R^2}-1\right)\tau_{\Theta\Theta}\right]\right\}\frac{r^2}{R^2}\tau_{\Theta\Theta}-p,\\
		&\sigma_{zz} =\mu \lambda_z^2-p.
		\end{aligned}
        \label{eq:sigmacomponents}
\end{equation}
These must satisfy  equilibrium  \eqref{eq:equilibrium equation}, which reduces to the component equation
\begin{equation}
	\frac{\mathrm{d}\sigma_{rr}}{\mathrm{d}r}+\frac{\sigma_{rr}-\sigma_{\theta\theta}}{r}=0
\end{equation}
that can be integrated  to obtain
\begin{equation}
    \sigma_{rr}=\int_a^r\frac{\sigma_{\theta\theta}-\sigma_{rr}}{r}\mathrm{d}r, \label{eq:sigmarr}
\end{equation}
where the lower limit $a$ is obtained from the traction-free condition on $r=a$. In view of the traction-free condition on $r=b$, we obtain
\begin{equation}
    \int_a^b\frac{\sigma_{\theta\theta}-\sigma_{rr}}{r}\mathrm{d}r=0, \label{eq:a(l)}
\end{equation}
where \eqref{eq:r with R} connects  $b$ and $R$ to  $a$. It follows from $\eqref{eq:sigmacomponents}_1$ and \eqref{eq:sigmarr} that the Lagrange multiplier $p$ is obtained by
\begin{equation}
%& p
%=\lambda_r \frac{\partial W}{\partial\lambda_r}- \sigma_{rr}=\mu \lambda_r^2+[1+\xi(I_5-\tr\tau)]\lambda_r^2\tau_{RR}-\sigma_{rr}
%&\\& 
p=\mu \frac{R^2}{r^2\lambda_z^2}+\left\{1+\xi \left[\left(\frac{R^2}{r^2 \lambda_z^2}-1\right)\tau_{RR}+\left(\frac{r^2}{R^2}-1\right)\tau_{\Theta\Theta}\right]\right\}\frac{R^2}{r^2\lambda_z^2}\tau_{RR}-\int_a^r\frac{\sigma_{\theta\theta}-\sigma_{rr}}{r}\mathrm{d}r.
\end{equation}

To evaluate \eqref{eq:sigmarr}, explicit expressions of the residual stress components $\tau_{RR}$, $\tau_{\Theta\Theta}$ that satisfy  equilibrium  \eqref{eq:43} and the boundary condition \eqref{eq:44} are needed. For the radial component we use 
\begin{equation}
\tau_{RR}=\zeta\left(R-A\right)\left(R-B\right),
\end{equation}
where $\zeta$ is a constant that defines the magnitude of the residual stress, see, for example,  \cite{dorfmann2021effect,merodio2016extension}. We note that $\tau_{RR}<0$ for $\zeta>0$. For the circumferential component $\tau_{\Theta\Theta}$ we use \eqref{eq:43} resulting in
\begin{equation}
\tau_{\Theta\Theta}=\zeta\left[3R^2-2(A+B)R+AB\right].
\end{equation}

It is convenient to take $\lambda_z$ as the loading parameter and use  \eqref{eq:a(l)} to obtain an implicit function connecting $a$ to $\lambda_z$, denoted $f(a,\lambda_z,\zeta)=0$. Once $a$ is identified, the deformed outer radius is given by \eqref{eq:r with R} and the primary deformation  is fully characterized.

To explore the influence of the residual stress on the primary deformation we first consider  the case where $\zeta=0$.  The constitutive equation \eqref{eq:nH} then specializes to the neo-Hookean model and the compression-induced primary deformation becomes homogeneous \citep{Wilkes1955AM,Dai2008ProceedingA,dai2015pitchfork}.

\begin{figure}[htp]
    \centering
        \includegraphics[width=0.45\textwidth]{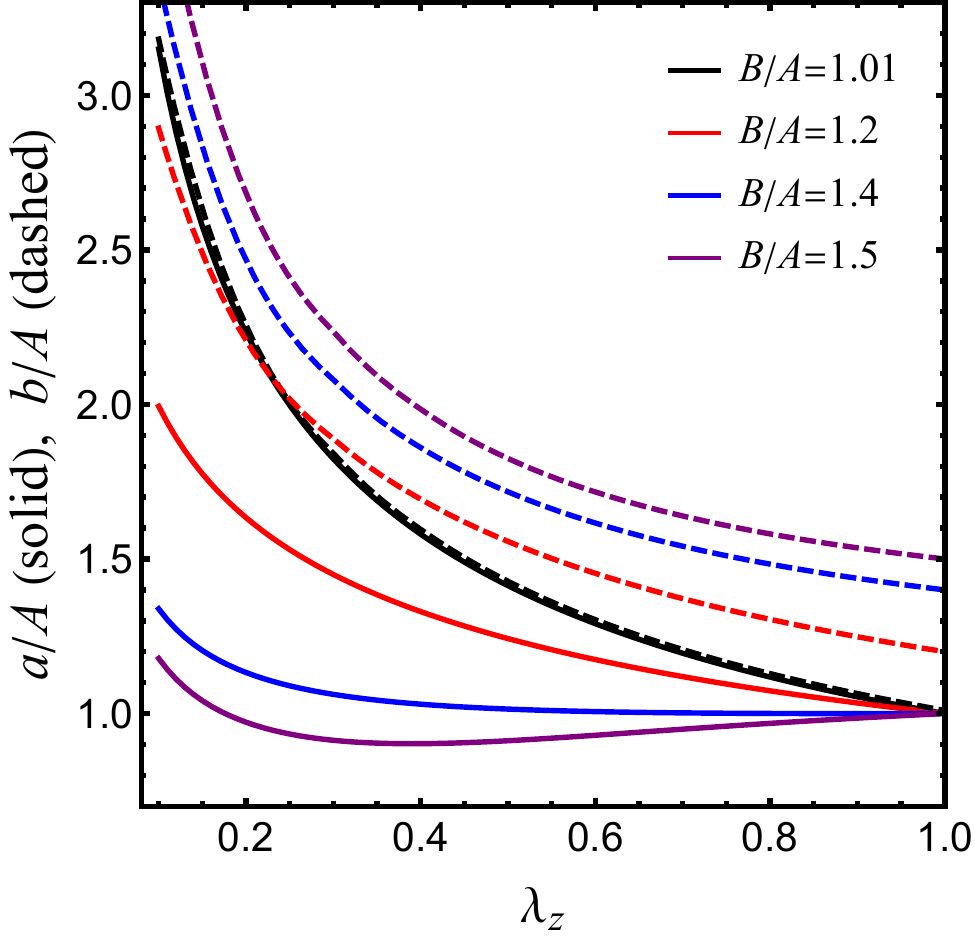}
        \includegraphics[width=0.45\textwidth]{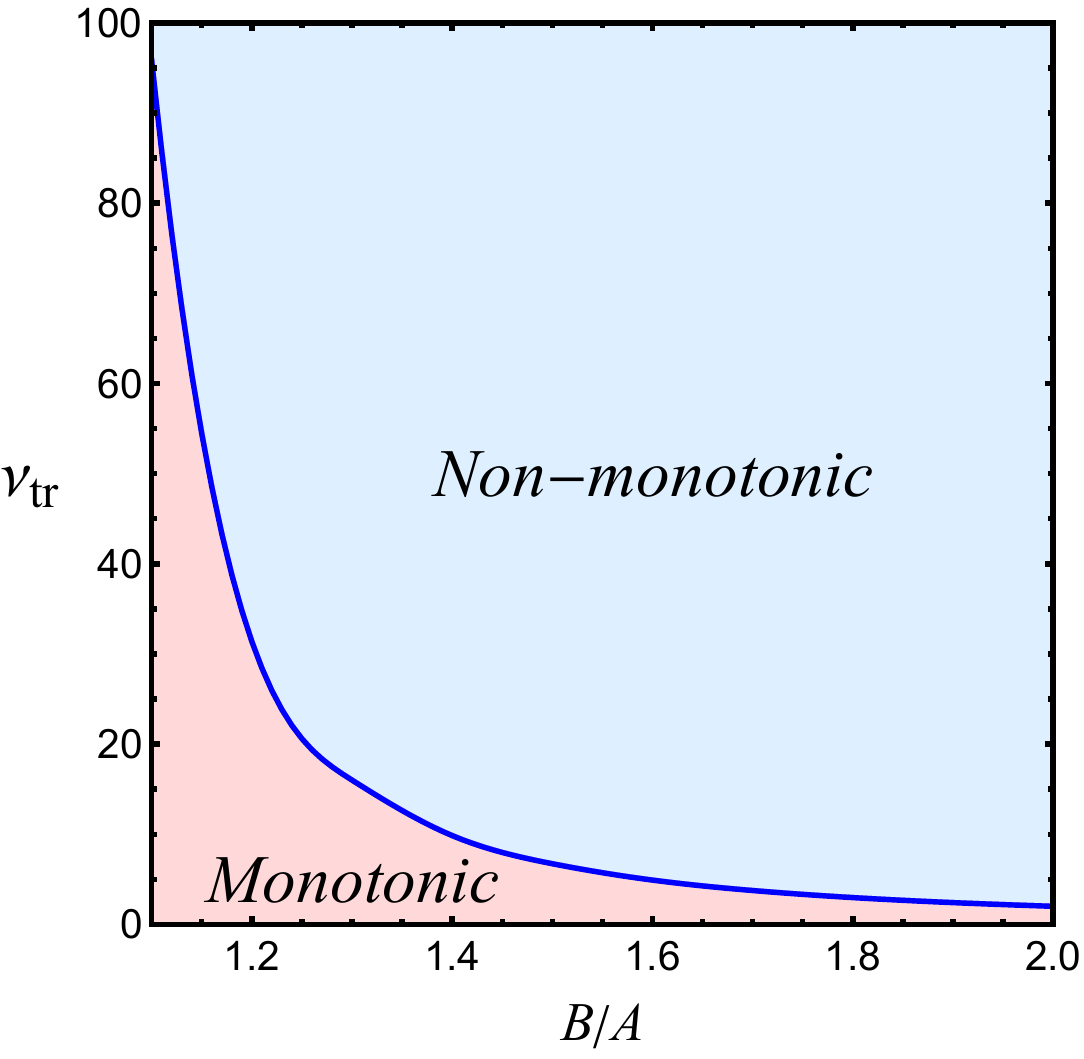}
    \caption{The left panel shows the dependence of the normalized  inner and outer radii on $\lambda_z$ for $\nu=10$ and $\kappa=0.5$. The one on the right the transition value $\nu_\mathrm{tr}$ as a function of $B/A$ for $\kappa=0.5$.}
    \label{fig:primary deformation}
\end{figure}

To investigate the effect of residual stress we introduce the dimensionless parameters $\kappa,\nu$  defined by 
\begin{equation}
    \kappa=\xi\mu,\quad \nu=\frac{\zeta A^2}{\mu},
    \label{eq:dim-kn}
\end{equation}
which  are used in the subsequent analysis.  In Figure \ref{fig:primary deformation} we show  the normalized  inner and outer radii $a/A$ and $b/A$ as a function of  the axial stretch $\lambda_z$ for the values  $\kappa=0.5$ and $\nu=10$. Note that the dashed curves  monotonically increase with the compression stretch  $\lambda_z$, indicating that the outer surface continuously expands.  The inner radius, for $B/A=1.01$ and $1.2$, shows a similar behavior. However, for larger values of $B/A$  a non-monotonic behavior is observed. Therefore, the inner surface first moves towards the center and then reverses direction for increasing compression. For a constant value of $B/A$, the transition point depends on the residual stress and can be obtained when the derivative of $a(\lambda_z)$ with respect to $\lambda_z$ vanishes at $\lambda_z=1$. With reference to the implicit function $f(a,\lambda_z,\zeta)=0$ we use
\begin{equation}
  \frac{\partial f(a,\lambda_z,\nu_\mathrm{tr})}{\partial\lambda_z}\Bigg|_{\lambda_z=1,a=A}=0,
\end{equation}
to obtain the transition value $\nu_\mathrm{tr}$. Representative values are listed in Table \ref{tab:transition} and illustrated as a function of $B/A$ in the right panel in Figure \ref{fig:primary deformation}. Hence, the residual stress plays a significant  role in  the primary deformation with its impact becoming more pronounced for increasing wall thickness. 
%\begin{figure}[htp]
%    \centering
%    {\subfigure[]{
%        \includegraphics[width=0.45\textwidth]{a-lambadz.pdf}
%        \label{fig:a(l)}}}
%    \quad
%    {\subfigure[]{
%        \includegraphics[width=0.45\textwidth]{vtr-B.pdf}
%        \label{fig:critical residual stress}}}
%    \caption{The left panel shows the dependence of the normalized  inner and outer radii on $\lambda_z$ for $\nu=10$ and $\kappa=0.5$. The one on the right the transition value $\nu_\mathrm{tr}$ as a function of $B/A$ for $\kappa=0.5$.}
%    \label{fig:primary deformation}
%\end{figure}

\begin{table}[h]
\centering
\caption{The transition residual stress parameter $\nu_\mathrm{tr}$ for different values of $B/A$.}
\begin{tabular}{ccccccccccc}
\toprule
$B/A$ & $1.1$ & $1.2$ & $1.3$ & $1.4$ & $1.5$ & 1.6 & 1.7 & 1.8 & 1.9 & 2.0 \\
\midrule
$\nu_\mathrm{tr}$ & 95.3710 & 31.2297 & 15.9761 & 9.8409 & 6.7217 & 4.9057 & 3.7501 & 2.9666 & 2.4096 & 1.9989 \\
\bottomrule
\end{tabular}
\label{tab:transition}
\end{table}

%%%%%%%%%%%%%%%%%%%%%%%%%%%%%%%%%%%%%%%%%%%%
\subsection{Bifurcation analysis}
%%%%%%%%%%%%%%%%%%%%%%%%%%%%%%%%%%%%%%%%%%%%

We now specialize the incremental theory summarized  in Section \ref{Incremental theory} to perform a buckling analysis of an axially compressed tube. We take  the incremental displacement vector $\mathbf u$ as
\begin{equation}
	\mathbf{u}=u\,\mathbf{e}_r+v\,\mathbf{e}_\theta+w\,\mathbf{e}_z,
\end{equation}
which  gives  the incremental displacement gradient \eqref{gradu} in matrix form
\begin{equation}
	\mathbf L=\left[\begin{array}{ccc}\dfrac{\partial u}{\partial r}&\dfrac{1}{r}\dfrac{\partial u}{\partial\theta}-\dfrac{v}{r}&\dfrac{\partial u}{\partial z}\\\\\dfrac{\partial v}{\partial r}&\dfrac{1}{r}\dfrac{\partial v}{\partial\theta}+\dfrac{u}{r}&\dfrac{\partial v}{\partial z}\\\\\dfrac{\partial w}{\partial r}&\dfrac{1}{r}\dfrac{\partial w}{\partial\theta}&\dfrac{\partial w}{\partial z}\end{array}\right],
\end{equation}
and the linearized incremental incompressibility condition \eqref{eq:imcompressible condition2} 
\begin{equation}
    \operatorname{tr}\mathbf{L}=\frac{\partial u}{\partial r}+\frac{1}{r}\frac{\partial v}{\partial\theta}+\frac{u}{r}+\frac{\partial w}{\partial z}=0.\label{eq:trL=0}
\end{equation}

The components of the incremental equation  \eqref{eq:expand increment equilibrium equation} become
\begin{equation}
	\begin{aligned}
		&\dot{T}_{011,1}+\dot{T}_{021,2}+\dot{T}_{031,3}+(\dot{T}_{011}-\dot{T}_{022})/r=0,\\
		&\dot{T}_{012,1}+\dot{T}_{022,2}+\dot{T}_{032,3}+(\dot{T}_{012}+\dot{T}_{021})/r=0, \\
		&\dot{T}_{013,1}+\dot{T}_{023,2}+\dot{T}_{033,3}+\dot{T}_{013}/r=0,
		\end{aligned}
        \label{eq:inc-eq}
\end{equation}
where we renamed the coordinate directions $r,\theta,z$ as $1,2,3$, respectively. From \eqref{eq:bc form} the traction-free boundary condition specializes to
\begin{equation}
	\dot{T}_{011}=0,\quad \dot{T}_{012}=0,\quad\dot{T}_{013}=0,\quad \textrm{on} ~ r=a,\, b.\label{eq:curved}
\end{equation}
Following \citet{Wilkes1955AM} and \citet{chen2017bifurcation}, we  assume that the two ends remain planar during  deformation and impose  sliding boundary conditions resulting in
\begin{equation}
    \frac{\partial w}{\partial \theta}=0,\quad \frac{\partial w}{\partial r}=0,\quad\dot{T}_{031}=0,\quad\dot{T}_{032}=0,\quad \textrm{on} ~ z=0,\, l.
    \label{eq:sliding}
\end{equation}

The Eulerian form of the  incremental stress components in \eqref{eq:increment version} have the explicit forms
\begin{equation}
	\begin{aligned}
		&\dot{T}_{011} =(\mathcal{A}_{01111}+p)L_{11}+\mathcal{A}_{01122}L_{22}+\mathcal{A}_{01133}L_{33}-\dot{p}, \\
		&\dot{T}_{022} = \mathcal A_{02211}L_{11}+(\mathcal A_{02222}+p)L_{22}+\mathcal A_{02233}L_{33}-\dot{p}, \\
		&\dot{T}_{033} = \mathcal{A}_{03311}L_{11}+\mathcal{A}_{03322}L_{22}+(\mathcal{A}_{03333}+p)L_{33}-\dot{p}, \\
		&\dot{T}_{012} = \mathcal{A}_{01212}L_{21}+(\mathcal{A}_{01221}+p)L_{12}, \\
		&\dot{T}_{021} = \mathcal{A}_{02121}L_{12}+(\mathcal{A}_{02112}+p)L_{21}, \\
		&\dot{T}_{013} = \mathcal{A}_{01313}L_{31}+(\mathcal{A}_{01331}+p)L_{13}, \\
		&\dot{T}_{031} = \mathcal{A}_{03131}L_{13}+(\mathcal{A}_{03113}+p)L_{31}, \\
		&\dot{T}_{023} =\mathcal{A}_{02323}L_{32}+(\mathcal{A}_{02332}+p)L_{23}, \\
		&\dot{T}_{032} = \mathcal{A}_{03232}L_{23}+(\mathcal{A}_{03223}+p)L_{32}.
		\end{aligned}\label{eq:expand stress}
\end{equation}

In addition to the dimensionless parameters  \eqref{eq:dim-kn}, we  introduce the normalized quantities 
\begin{equation}
    \begin{aligned}
       &\hat{r}=\frac{r}{A},\quad\hat{z}=\frac{z}{A},\quad\hat{u}=\frac{u}{A},\quad\hat{w}=\frac{w}{A},\quad\hat{a}=\frac{a}{A},\quad\hat{b}=\frac{b}{A},\\&
       \hat{l}=\frac{l}{A},\quad\gamma=\frac{B}{L}, \quad \hat{\dot{T}}_{0ij}=\frac{\dot{T}_{0ij}}{\mu},\quad\hat{p}=\frac{p}{\mu},\quad\hat{\dot{p}}=\frac{\dot{p}}{\mu},\quad\hat{\mathcal{A}}_{0jilk}=\frac{\mathcal{A}_{0jilk}}{\mu}.
    \end{aligned}
\end{equation}
For clarity,    the hats on these  variables will be dropped in the subsequent analysis.

To derive the bifurcation condition, we employ  Stroh's method \citep{stroh1962steady} and look for periodic solutions of the form  
\begin{equation}
    \begin{aligned}
        &u= U(r)\cos m\theta\cos\alpha z,\quad v= V(r)\sin m\theta\cos\alpha z,\quad w= W(r)\cos m\theta\sin\alpha z, \\&
        r\dot{T}_{011} = S_{11}(r)\cos m\theta\cos\alpha z, \quad r\dot{T}_{012} = S_{12}(r)\sin m\theta\cos\alpha z, \quad r\dot{T}_{013} = S_{13}(r)\cos m\theta\sin\alpha z, \\
    \end{aligned}
    \label{eq:per-s}
\end{equation}
where $m$ is the circumferential  and $\alpha$ the axial wave number. The end conditions \eqref{eq:sliding} result in
\begin{equation}
	\alpha=\frac{n\pi}l=\frac{n \pi}{\lambda_z L/A}=\frac{\gamma n\pi}{\lambda_zB/A},\quad n=1,2,3,\cdots,
    %=\frac{n \pi}{\lambda_z L}=\frac{\gamma n\pi}{B\lambda_z}
    \label{eq:alpha}
\end{equation}
where $n$ denotes the normalized axial wave number. It will be shown that this selection connects $\gamma$ and $n$ and  facilitates the bifurcation analysis by reducing the number of free parameters.

To write the equations in Stroh form,  we define the displacement-traction vector $\boldsymbol{\eta}$ as 
\begin{equation}
    \left.\boldsymbol\eta(r)=\left[\begin{array}{c}\mathbf{U}(r)\\r\mathbf{S}(r)\end{array}\right.\right]\quad\textrm{where}\quad\left\{\begin{array}{l}\mathbf{U}(r)=\left[U(r),V(r),W(r)\right]^\mathrm{T},\\\mathbf{S}(r)=\left[S_{11}(r),S_{12}(r),S_{13}(r)\right]^\mathrm{T}.\end{array}\right.\label{eq:displacement-traction vector}
\end{equation}
We also take  $\dot{p}=P(r)\cos m\theta\cos\alpha z$ and use of $\eqref{eq:per-s}_4$ to obtain
\begin{equation}
    P(r)=\frac{1}{r}\left(\mathcal{A}_{1122}(U(r)+mV(r))-S_{11}(r)+r\left((\mathcal{A}_{1122}+p)U^\prime(r)-\alpha\mathcal{A}_{1133}W(r)\right)\right).
\end{equation}
In addition, using $\eqref{eq:inc-eq}_{1,2}$ and $\eqref{eq:expand stress}_{1,4,6}$, we find that
\begin{equation}
	\frac{\mathrm{d}\boldsymbol{\eta}(r)}{\mathrm{d}r}=\frac{1}{r}\mathbf{G}(r)\boldsymbol{\eta}(r),\label{eq:rewritten matrix}
\end{equation}
which is known as the Stroh form of the governing equations.  The Stroh matrix $\mathbf{G}(r)$ has the form
\begin{equation}
	\mathbf{G}(r)=
    \begin{bmatrix}
    \mathbf{G}_1&\mathbf{G}_2\\\mathbf{G}_3& \mathbf{G}_4,
    \end{bmatrix}.
\end{equation}
where $\mathbf G_i, i=\{1,2,3\}$ are $3\times 3$ matrices, with $\mathbf G_2$ and $\mathbf G_3$ symmetric. In component form we have
\begin{align}
       \nonumber &\mathbf{G}_{1} = \begin{bmatrix}
            -1 & -m & -\alpha r \\
            \dfrac{m(p+\mathcal{A}_{1221})}{\mathcal{A}_{1212}} & \dfrac{p+\mathcal{A}_{1221}}{\mathcal{A}_{1212}} & 0 \\
            \dfrac{\alpha (p+\mathcal{A}_{1331}) r}{\mathcal{A}_{1313}} & 0 & 0
        \end{bmatrix},\quad
        \mathbf{G}_{2} = \begin{bmatrix}
            0 & 0 & 0 \\
            0 & \dfrac{1}{\mathcal{A}_{1212}} & 0 \\
            0 & 0 & \dfrac{1}{\mathcal{A}_{1313}}
        \end{bmatrix},\\&
        \mathbf{G}_{3} = \begin{bmatrix}
            g_{11} & g_{12} & g_{13} \\
            g_{12} & g_{22} & g_{23} \\
            g_{13} & g_{23} & g_{33}
        \end{bmatrix}, \quad \mathbf{G}_4=-\mathbf{G}_{1}^{\mathrm{T}},
        \label{eq:stroh matrix}
\end{align}
where the components of $\mathrm{G}_3$ have the expressions 
\begin{equation}
    \begin{aligned}
          g_{11}=& ~\mathcal{A}_{1111}-2\mathcal{A}_{1122}+\mathcal{A}_{2222}+\mathcal{A}_{2121} m^2-\frac{m^2 \left(\mathcal{A}_{1221}+p\right)^2}{\mathcal{A}_{1212}} -\frac{\alpha ^2 r^2 \left(\mathcal{A}_{1331}+p\right)^2}{\mathcal{A}_{1313}}+2 p+\alpha ^2  r^2 \mathcal{A}_{3131},\\
        g_{12}=& ~m \left(\mathcal{A}_{1111}-2\mathcal{A}_{1122}+\mathcal{A}_{2121}+\mathcal{A}_{2222}-\frac{\left(\mathcal{A}_{1221}+p\right)^2}{\mathcal{A}_{1212}}+2 p\right),\\
        g_{13}=& ~\alpha  r \left(\mathcal{A}_{1111}-\mathcal{A}_{1133}-\mathcal{A}_{2211}+\mathcal{A}_{2233}+p\right),\\
        g_{22}=& ~\mathcal{A}_{2121}+\mathcal{A}_{1111} m^2-2\mathcal{A}_{1122} m^2+\mathcal{A}_{2222} m^2+2 m^2 p-\frac{\left(\mathcal{A}_{1221}+p\right)^2}{\mathcal{A}_{1212}}+\alpha ^2  r^2 \mathcal{A}_{3232},\\
        g_{23}=& ~\alpha r \left(\mathcal{A}_{1111}-\mathcal{A}_{1133}-\mathcal{A}_{2211}+\mathcal{A}_{2233}+\mathcal{A}_{3232}+2 p\right) m,\\
        g_{33}=& ~\mathcal{A}_{2323} m^2+\alpha ^2 r^2 \left(\mathcal{A}_{1111}-2\mathcal{A}_{1133}+\mathcal{A}_{3333}+2 p\right).
    \end{aligned}
    \label{eq:G3}
\end{equation}

Without loss of generality, we assume that  \eqref{eq:rewritten matrix} admits six independent solutions  $\boldsymbol\eta_i(r), i=\{1,\dots,6\}$, which suggests to write  a general solution as 
\begin{equation}
    \boldsymbol\eta(r)=\boldsymbol\Gamma(r)\mathbf{d},
\end{equation}
where 
\begin{equation}
    \boldsymbol\Gamma(r)=[\boldsymbol\eta_1,\boldsymbol\eta_2,\boldsymbol\eta_3,\boldsymbol\eta_4,\boldsymbol\eta_5,\boldsymbol\eta_6],\quad \mathbf{d}=[C_1,C_2,C_3,C_4,C_5,C_6]^\mathrm{T},
\end{equation}
with $C_i$ $\{i=1,\dots,6\}$ being arbitrary constants. For any $a\leqslant r_k\leqslant b$, we define 
\begin{equation}
    \left.\mathbf{M}(r,r_k)=\boldsymbol\Gamma(r)\boldsymbol\Gamma^{-1}(r_k)=\left(\begin{array}{cc}\mathbf{M}_1(r,r_k)&\mathbf{M}_2(r,r_k)\\\mathbf{M}_3(r,r_k)&\mathbf{M}_4(r,r_k)\end{array}\right.\right),\quad \mathbf{D}=\boldsymbol\Gamma(r_k)\mathbf{d},
\end{equation} 
and rewrite the  general solution as 
\begin{equation}
    \boldsymbol\eta(r)=\mathbf{M}(r,r_k)\mathbf{D}.
\end{equation}
It is easy to show that  $\mathbf{M}\left(r_k,r_k\right)$ is the $6\times6$ identity matrix.  Then it follows from \eqref{eq:rewritten matrix} that 
\begin{equation}
    \frac{\mathrm{d}\mathbf{M}\left(r,r_k\right)}{\mathrm{d}r}=\frac{1}{r}\mathbf{G}(r)\mathbf{M}\left(r,r_k\right).
\end{equation}

We use  the surface impedance matrix method and assume that there exists a conditional impedance matrix $\mathbf{z}(r)$ such that 
\begin{equation}
    r\mathbf{S}(r)=\mathbf{z}(r)\mathbf{u}(r).\label{eq:impedance matrix}
\end{equation}
For details we refer to, for example,  \citep{shuvalov2003frobenius,shuvalov2003sextic,norris2010wave}. 
On substituting  \eqref{eq:displacement-traction vector} and \eqref{eq:impedance matrix} into  \eqref{eq:rewritten matrix} we obtain the  Riccati equation
\begin{equation}
    \frac{\mathrm{d}}{\mathrm{d}r}\mathbf{z}=\frac{1}{r}\begin{pmatrix}\mathbf{G}_3-\mathbf{G}_1^\mathrm{T}\mathbf{z}-\mathbf{z}\mathbf{G}_1-\mathbf{z}\mathbf{G}_2\mathbf{z}\end{pmatrix}.\label{eq:riccati}
\end{equation}

Taking $r_k=a$ and making use of the traction-free condition on $r=a$ gives 
\begin{equation}
   \mathbf{z}=\mathbf{M}_3\mathbf{M}_1^{-1}.
\end{equation}
Since $\mathbf{M}_3(a)=0$ we  obtain $\mathbf{z}(a)=\mathbf{0}$, which  yields the initial condition to numerically integrate \eqref{eq:riccati} from $r=a$ to $r=b$. Traction-free boundary conditions on $r=b$ result in $\mathbf{z}(b)\mathbf{u}(b)=\mathbf{0}$ and  in the bifurcation condition
\begin{equation}
    \det[\mathbf{z}(b)]=0,\label{eq:bifurcation condition}
\end{equation}
which depends on the axial stretch $\lambda_z$,  the residual stress parameter $\nu$,  the material parameter $\kappa$,  the slenderness ratio $\gamma=B/L$, the mode numbers $n$, $m$, and the wall thickness ratio $B/A$. We use the software package $\mathit{Mathematica}$ \citep{wolfram2020mathematica} to solve the bifurcation condition \eqref{eq:bifurcation condition} numerically.

%%%%%%%%%%%%%%%%%%%%%%%%%%%%%%%%%%%%%%%%%%%%%%%%%%%%%%%
\section{Parametric study of the buckling behavior}\label{Parametric study}
%%%%%%%%%%%%%%%%%%%%%%%%%%%%%%%%%%%%%%%%%%%%%%%%%%%%%%%

In this section we consider tubes with  slenderness ratio $\gamma$,  with thickness ratio $B/A$ and with residual stress and stiffness parameters $\nu$ and $\kappa$ to evaluate the effect of the residual stress on the buckling behavior. When  $\mathcal{A}_{1212} = 0$ or $\mathcal{A}_{1313} = 0$, special attention is necessary to treat the singularity when integrating  \eqref{eq:rewritten matrix} from $a<r<b$. With no residual stress $\mathcal{A}_{1212} = \mathcal{A}_{1313} = \lambda_z^{-1}$ and the singularity does not occur. However, as discussed in \citet{liu2024localized}, the residual stress  generates at least one regular singular point. For a  tube, this can be excluded by limiting the wall thickness. We therefore restrict attention to a moderately thick-walled tube with $1<B/A\leqslant1.4$.

\begin{figure}[!h]
    \centering
        \includegraphics[width=0.46\textwidth]{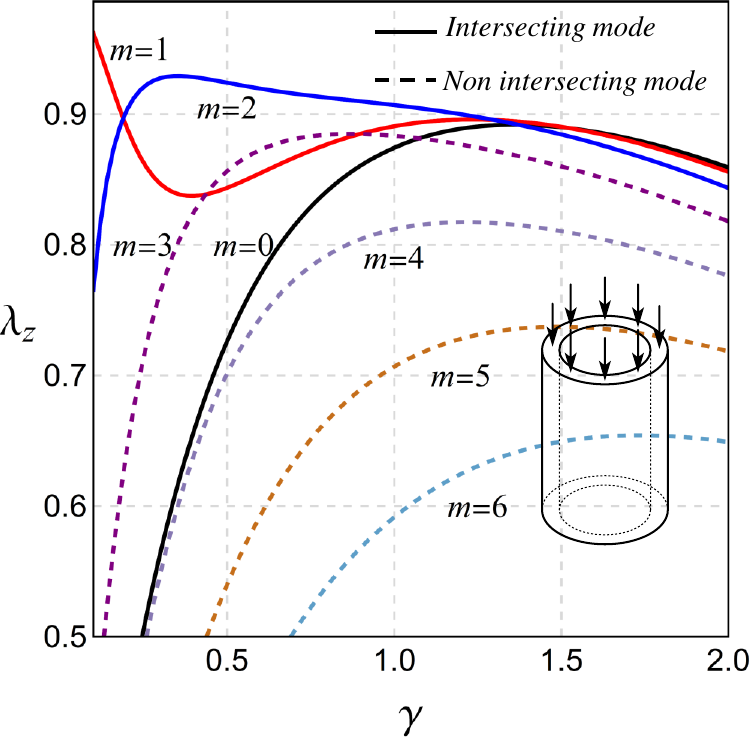}
        \includegraphics[width=0.46\textwidth]{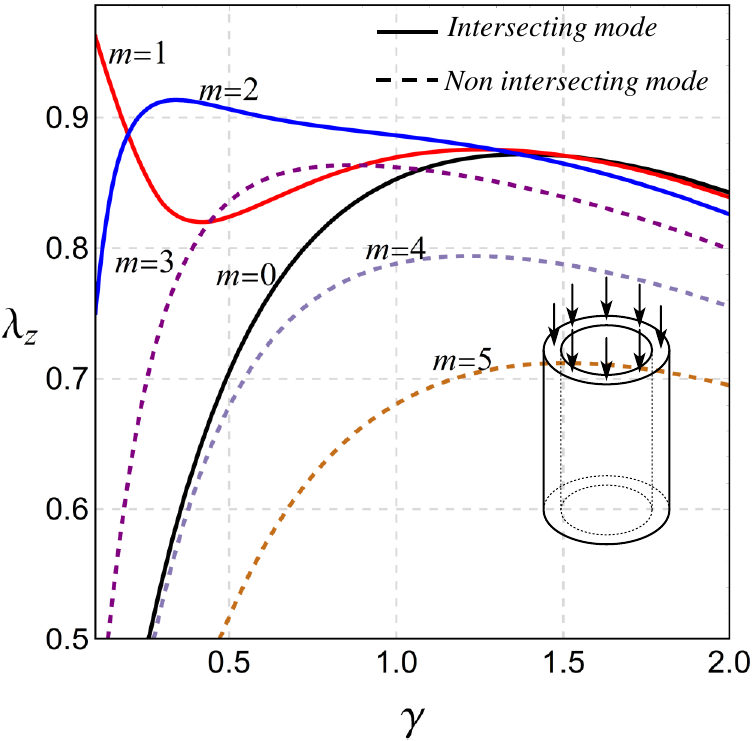}
    \caption{Bifurcation curves obtained from  the  condition \eqref{eq:bifurcation condition} for various circumferential wave numbers. The panel on the left is for  $\kappa=0.5,~B/A=1.2,~\nu = 5$. The results on the right are for $\kappa=0.5,~B/A=1.2,~\nu = -5$.}
    \label{fig:buckling diagram1}
\end{figure}

Note that the slenderness parameter $\gamma$ and the axial wave number $n$   appear together in the bifurcation condition, whose dependence on $\gamma$ is through $\alpha$, see \eqref{eq:alpha}. Therefore, it is convenient to treat $\gamma n$ as a single parameter setting $n = 1$. Then,  in Figure \ref{fig:buckling diagram1} we illustrate the bifurcation curves for the dimensionless parameters $\kappa=0.5$ and $B/A=1.2$ and for different circumferential modes as a function of $\gamma$. The left and right panels illustrate the responses for  $\nu=5$ and $\nu=-5$, respectively. For fixed $\gamma$, the  value of $\lambda_z$ closed to unity gives the critical stretch. Following \cite{springhetti2023buckling}, we depict the curves that provide the critical stretch as solid lines, known as intersecting modes. The results indicate that for increasing  $\gamma$ the critical  wave number changes from $m = 1$ to $m = 2$ and then back to $m = 1$. For further increase in $\gamma$ the barreling mode $m = 0$ occurs first.  With no residual stress, the classical results of compressed tubes reported by, for example,  \citet{goriely2008nonlinear, springhetti2023buckling} are recovered.

\begin{figure}[!h]
    \centering
        \includegraphics[width=0.47\textwidth]{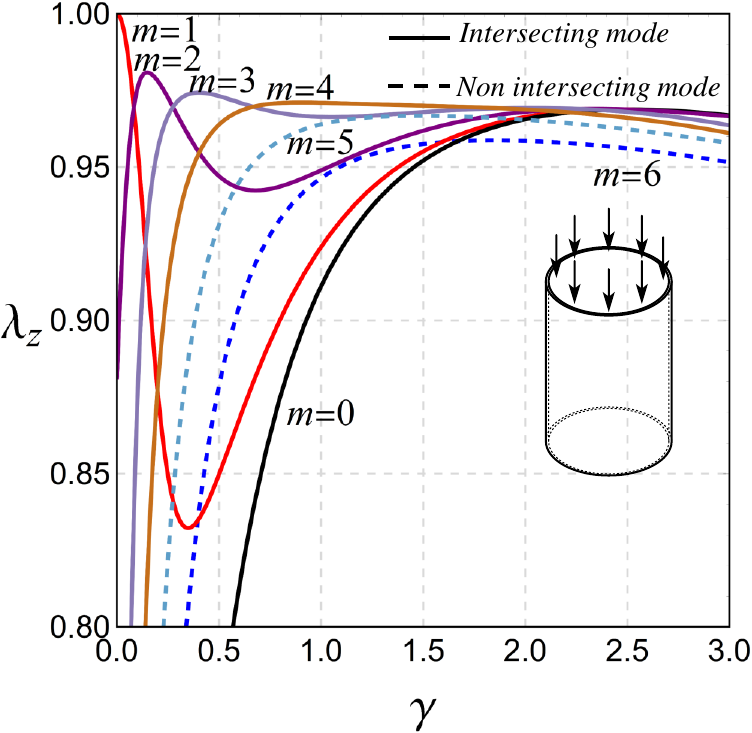}
        \includegraphics[width=0.46\textwidth]{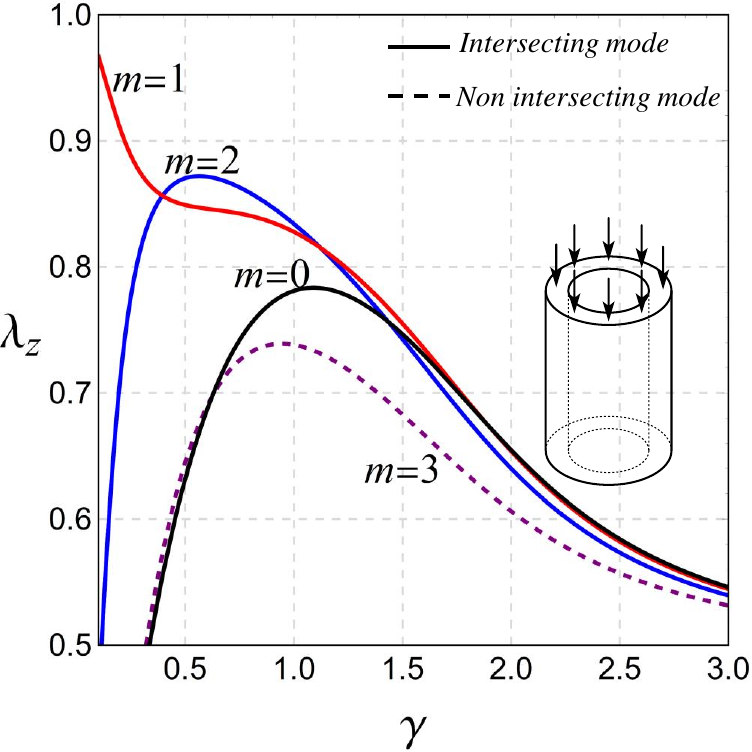}
    \caption{The critical stretch $\lambda_z$ for different circumferential wave numbers. The results on the left  are for $\kappa=0.5,~B/A=1.05,~\nu = 5$, the ones on the right for $\kappa=0.5,~B/A=1.4,~\nu = 5$.}
    \label{fig:buckling diagram2}
\end{figure}

In Figure \ref{fig:buckling diagram2} we illustrate the bifurcation parameter $\lambda_z$ as a function of $\gamma$ for  two values of   $B/A$   when $\kappa=0.5$ and $\nu=5$. The curves  in the left panel are obtained for $B/A = 1.05$ and  show that the critical mode changes  from $m=0$ to $m=4$, depending on the slenderness ratio $\gamma$. The results on the right correspond to  $B/A=1.4$ and again show that the critical mode numbers depend on $\gamma$ and are $m=0,1,2$.  Comparing  to the results in  Figure \ref{fig:buckling diagram1}, we conclude  that  increasing wall-thickness delays the onset of instability.  We also find the dependence of the critical stretch on the slenderness ratio $\gamma$ complex due to mode transitions and, even for a fixed mode,  the behavior of $\lambda_z$ as a function of $\gamma$ is  non-monotonic. When $\gamma\rightarrow0$, the tube approaches the membrane limit and  Euler buckling $m=1$ occurs  in all cases. Conversely, as $\gamma\rightarrow\infty$, the critical stretch is attained when $m=0$, similar to  the results given in \citet{Wilkes1955AM}. 

\begin{figure}[!h]
    \centering
        \includegraphics[width=0.46\textwidth]{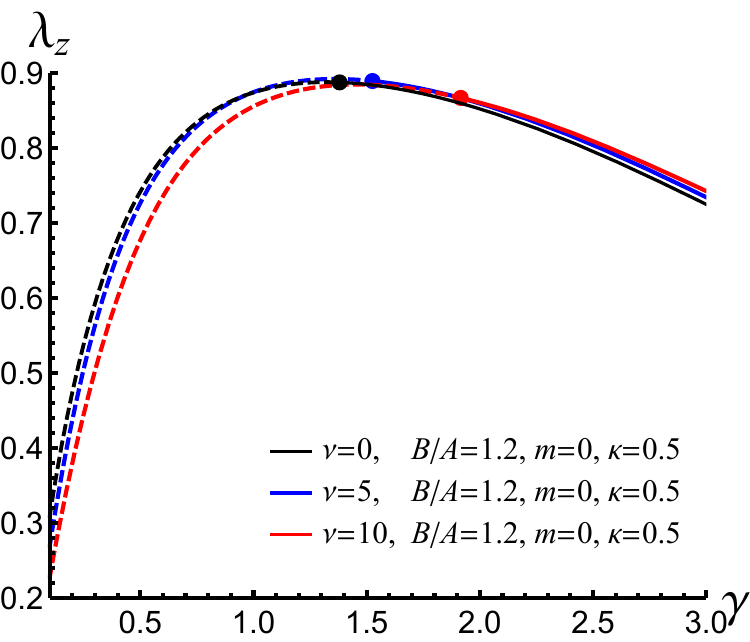}
        \includegraphics[width=0.46\textwidth]{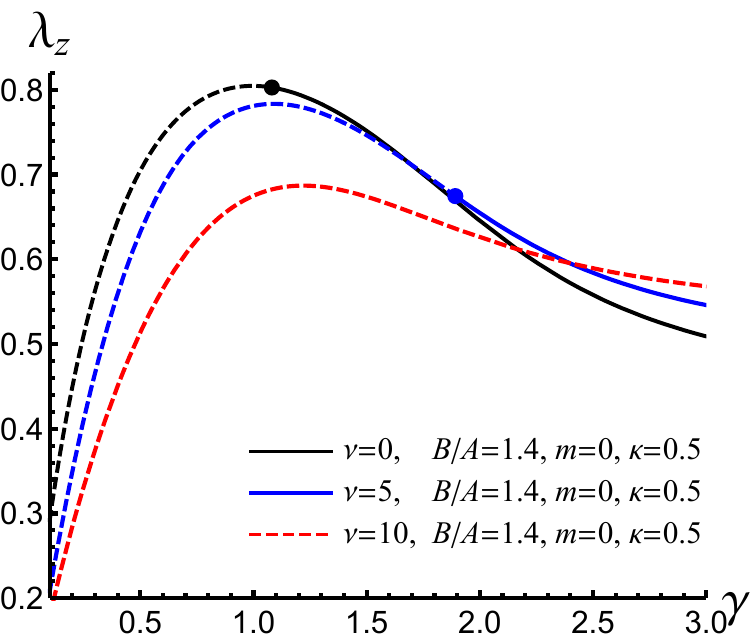}
    \caption{The critical stretch $\lambda_z$  for barreling to occur as a function of the slenderness parameter $\gamma$ for   $\nu = 0, 5, 10$. The results shown in the left panel are for $\kappa=0.5$ and $B/A = 1.2$, those  on the right  for  $\kappa=0.5$ and  $B/A = 1.4$.}
    \label{fig:axisymmetric mode(lambda-gamma)}
\end{figure}

We now focus on the range of $\gamma$, where the barreling mode $m=0$ occurs  \citep{chau1995buckling}. In Figure \ref{fig:axisymmetric mode(lambda-gamma)}  we illustrate the critical stretch $\lambda_z$ as a function of $\gamma$ for values of the dimensionless residual stress parameter  $\nu = 0, 5, 10$. The dashed curves represent the regions where Euler buckling occurs and  the dots indicate the transition  to barreling . The curves in the left panel show that for a tube with $B/A=1.2$ the influence of the residual stress  is essentially identical for all  considered values of $\nu$. The bifurcation curves on the right    show that  for a tube with $B/A=1.4$ the residual stress leads to a delay in the emergence of the barreling mode, i.e., more compression is required. For $\nu=10$, for example,  the barreling mode is entirely suppressed, suggesting that a residual stress can effectively eliminate some instability modes. 

\begin{figure}[!h]
\centering
        \includegraphics[width=0.7\textwidth]{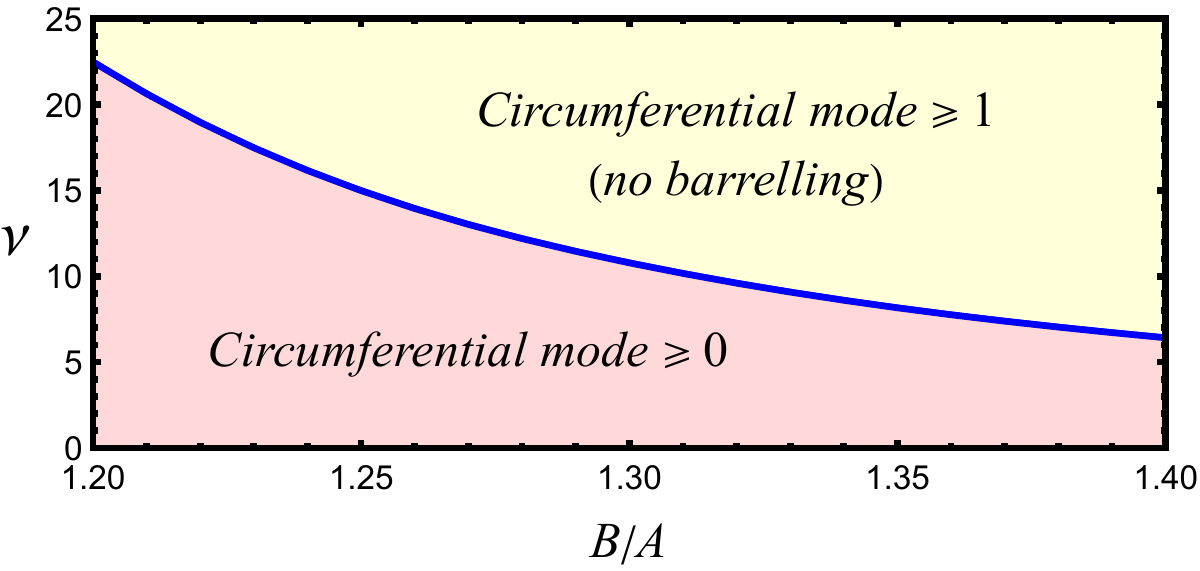}
        \caption{Phase diagram of different circumferential modes in the  $\nu-B/A$ plane when $\kappa=0.5$.}
        \label{fig:mode=0 vanished}
\end{figure}

In  \cite{springhetti2023buckling} it is shown that  barreling instability  can occur in the absence of residual stress. It is therefore of interest to determine  the value of the dimensionless parameter $\nu$  necessary  to  suppress this mode of instability, see Figure \ref{fig:axisymmetric mode(lambda-gamma)}. We take $\kappa=0.5$,  $B/A=1.4$,  select a ratio $\gamma$ and  simultaneously solve the bifurcation condition for  $\lambda_z$ and $\nu$
\begin{equation}
    \det[\mathbf{z}(b)]\Big|_{m=0}=0,\quad    \det[\mathbf{z}(b)]\Big|_{m=1}=0. \label{eq:bifurcation condition-2}
\end{equation}

By varying $\gamma$ we obtain a  curve in the $\nu -\gamma$ plane representing the transition from the barreling mode $m=0$ to the Euler buckling mode $m=1$. This curve is non-monotonic and its maximum occurs at $\nu \approx 6.41$. Hence, for $\nu > 6.41$ the barreling mode is completely suppressed consistent with the results depicted in the right panel of Figure \ref{fig:axisymmetric mode(lambda-gamma)}. Results obtained by varying the ratio $B/A$ are shown in the phase diagram in Figure \ref{fig:mode=0 vanished}.

\begin{figure}[!h]
    \centering
        \includegraphics[width=0.45\textwidth]{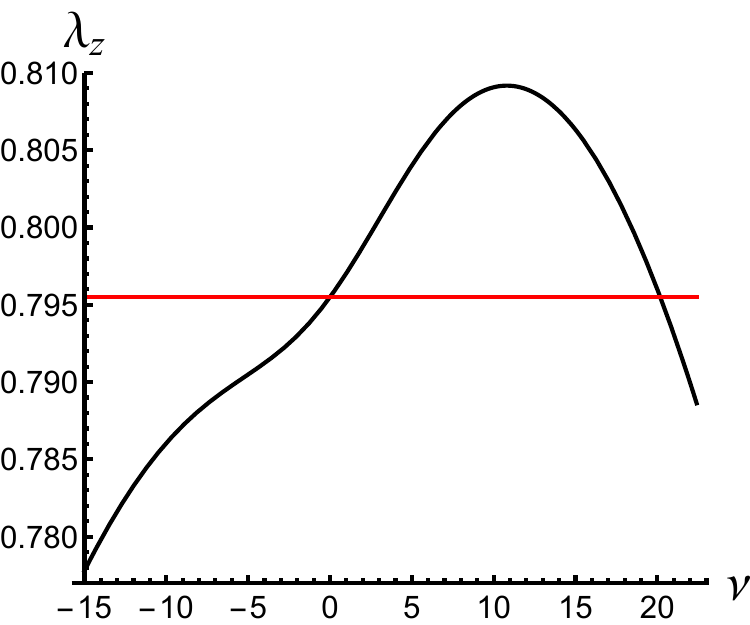}
        \includegraphics[width=0.45\textwidth]{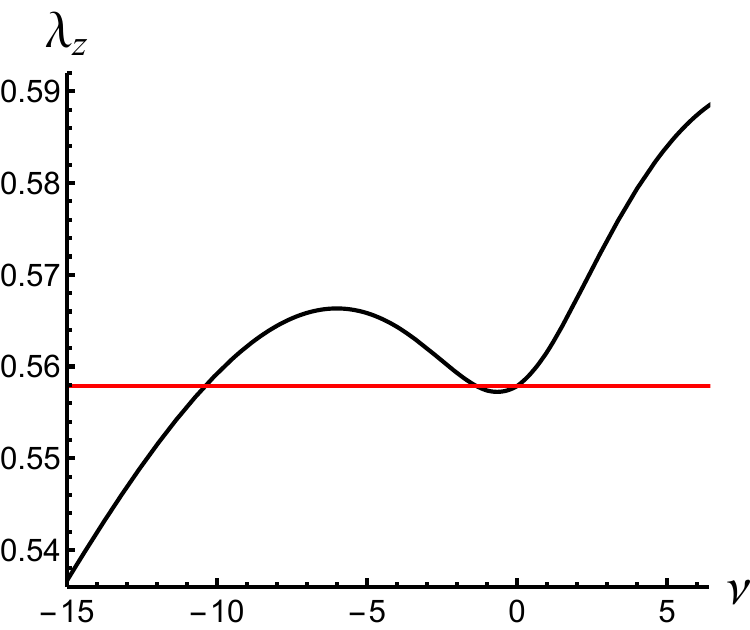}
    \caption{Dependence of the critical stretch of the barreling mode $m=0$ on the residual stress $\nu$, for  $\kappa=0.5$ and $\gamma=2.5$. The results  on the left are obtained when $B/A=1.2$, the ones on the right when $B/A=1.4$.}
    \label{fig:m0cr}
\end{figure}

In  Figure \ref{fig:m0cr}  we show the critical stretch $\lambda_z$ as a function of the dimensionless parameter $\nu$ for the barreling mode $m=0$, for $\kappa = 0.5$ and  $\gamma = 2.5$. The results shown on the left are obtained when $B/A=1.2$, the ones on the right when $B/A=1.4$.

The horizontal line in each panel is used for reference and marks the critical stretch for a compressed tube with no residual stress.  The results on the left are for $-15\leqslant \nu \leqslant 22.49$ and show that  negative values delay the onset of instability while positive values render the tube more unstable. For $\nu\approx20.182$ the residual stress has no effect, with the critical stretch being identical to the tube with no residual stress. For $\nu>20.182$ a delay in the onset of instability occurs.

The panel on the right shows the  stability behavior for $-15\leqslant \nu \leqslant 6.41$, where  the upper limit represents the threshold beyond which the barreling mode no longer exists. The results show that the horizontal line and the curve intersect when  $\nu \approx -10.398$ and $\nu \approx -1.353$, indicating that the residual stress has no effect on the tube stability. Bifurcation is delayed for $-15\leqslant \nu \leqslant -10.398$ and $-1.353\leqslant \nu \leqslant 0$. Outside these ranges the residual stress destabilizes the tube. 

\begin{figure}[!h]
    \centering
        \includegraphics[width=0.45\textwidth]{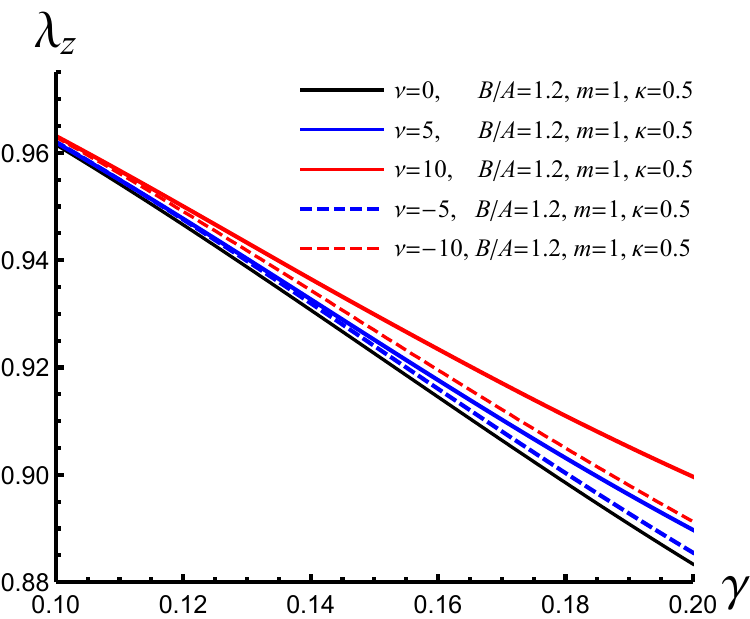}
        \includegraphics[width=0.45\textwidth]{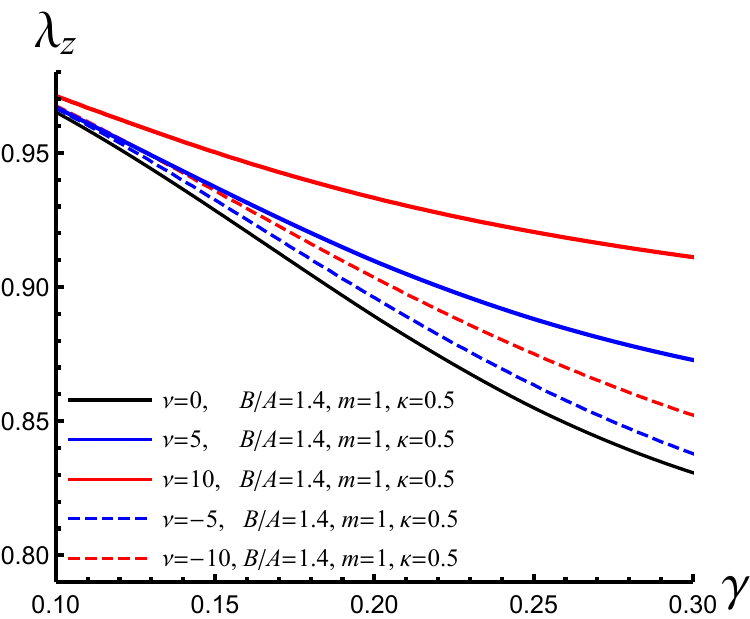}
    \caption{Bifurcation curves for the buckling mode $m=1$ at different magnitudes of the residual stress $\nu$. The parameters are set as (a) $\kappa=0.5$, $B/A = 1.2$ and (b) $\kappa=0.5$, $B/A = 1.4$.}
    \label{fig:asymmetric solution}
\end{figure}

We now investigate the influence of the residual stress and slenderness ratio $\gamma=B/L$ on the Euler buckling mode $m = 1$  \citep{goriely2008nonlinear, zhou2023three}. From  Figures \ref{fig:buckling diagram1} and \ref{fig:buckling diagram2} we find that this instability  occurs for  small values of $\gamma$, hence we  focus on this region. Figure \ref{fig:asymmetric solution} shows the bifurcation curves for $m = 1$, $\kappa = 0.5$ and  $\nu=0, \pm 5,\pm 10$.  The curves in the left panel are obtained when $B/A=1.2$ and the ones on the right are for $B/A = 1.4$. The solid, black curves correspond to the bifurcation with no residual stress, which in all cases lie below the others. Therefore, the residual stress has a destabilizing effect. We also find that an increase in $\gamma$ due to an increase in the outer radius or a reduction in the total length $L$ stabilizes the tube. Comparing the left and right images we note that an increase in the wall-thickness amplifies the effect of the residual stress, which has negligible influence in very thin tubes. 

\begin{figure}[!h]
    \centering
        \includegraphics[width=0.45\textwidth]{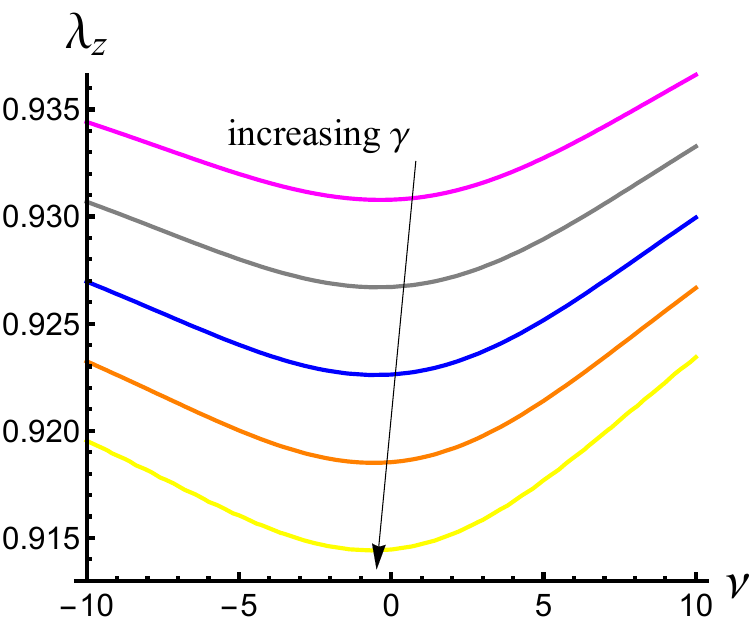}
        \includegraphics[width=0.45\textwidth]{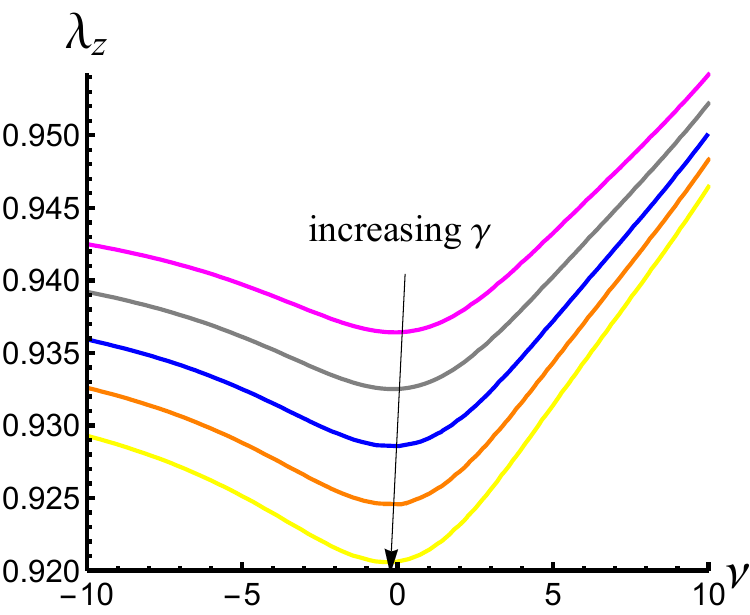}
    \caption{The critical stretch of the buckling mode $m = 1$ as a function of the dimensionless parameter  $\nu$ with $\kappa = 0.5$ and $\gamma$ varying from $0.14$ to $0.16$ with increments of $0.005$. The results shown on the left are obtained when $B/A=1.2$, the ones on the right when $B/A=1.4$.}
    \label{fig:m1cr}
\end{figure}

In Figure \ref{fig:m1cr} we depict the dependence of the critical stretch $\lambda_z$ on the dimensionless residual stress parameter  $\nu$ for the fixed value $\kappa = 0.5$. The curves from top to bottom correspond to slenderness ratios $\gamma$ from  $0.14$ to $0.16$ with increments of  $0.005$. Each curve has a minimum where $\nu \approx 0$ indicating that both positive and negative values of $\nu$ destabilize the tube.

\begin{figure}[!h]
    \centering
        \includegraphics[width=0.45\textwidth]{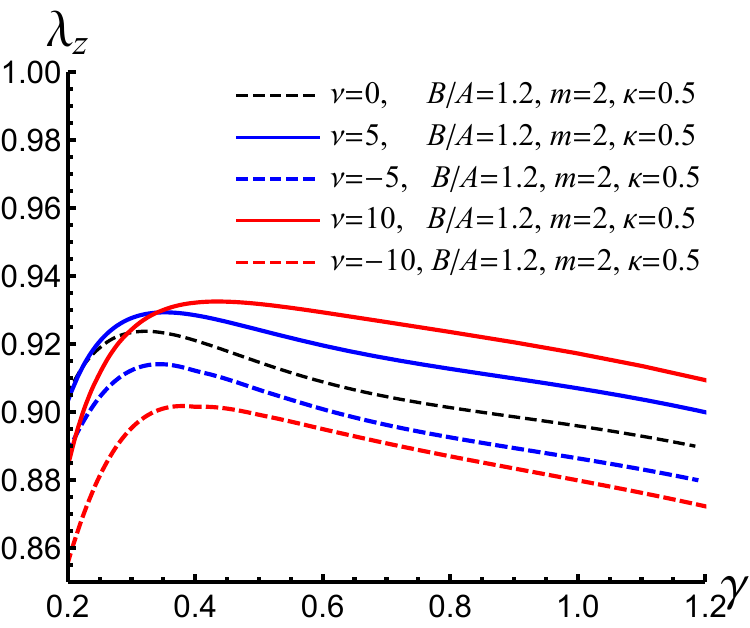}
        \includegraphics[width=0.45\textwidth]{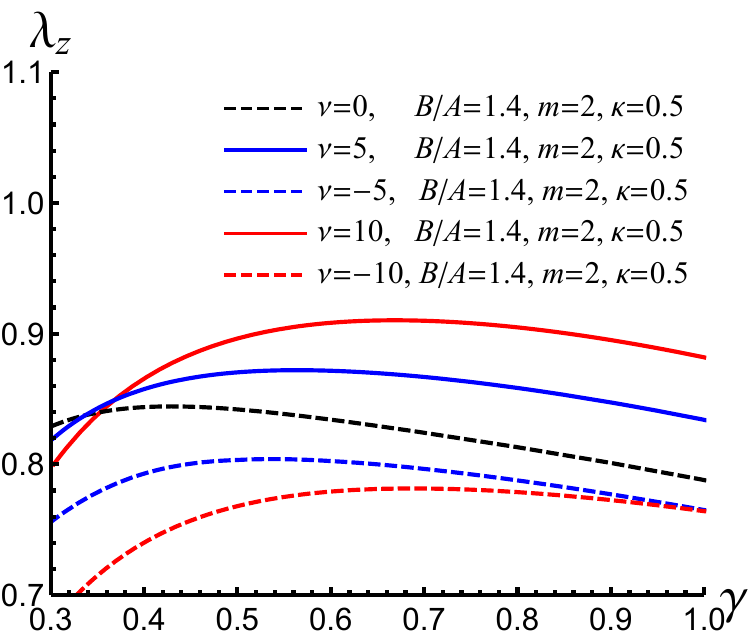}
    \caption{Bifurcation curves for mode $m=2$, $\kappa=0.5$ and $\nu=0,\pm5,\pm10$. The results shown on the left are for $B/A=1.2$, the ones on the right for $B/A=1.4$.}
    \label{fig:m=2}
\end{figure}

In figures \ref{fig:buckling diagram1} and \ref{fig:buckling diagram2} we have shown that the  buckling  mode transitions from $m=1$ to $m=2$ as $\gamma$ increases. Here we investigate how the  residual stress effects mode $m=2$ and illustrate the results in Figure \ref{fig:m=2}. We plot the bifurcation curves for $m=2$, $\kappa=0.5$  and $\nu=0,\pm 5,\pm 10$,  on the left for $B/A=1.2$ and on the right for $B/A=1.4$. We note that the critical stretch $\lambda_z$ exhibits a non-monotonic dependence on $\gamma$ and  that an increase in the wall-thickness amplifies the effect of the residual stress. 

\begin{figure}[!h]
    \centering
        \includegraphics[width=0.45\textwidth]{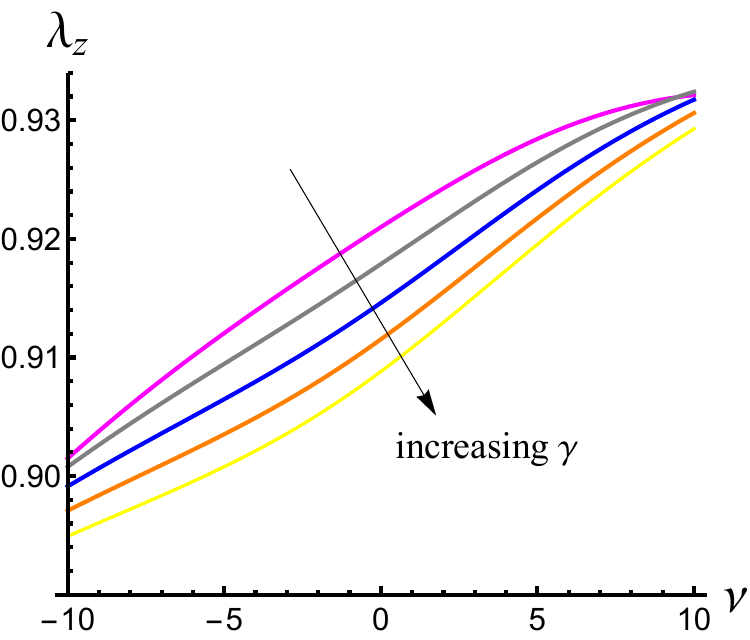}
        \includegraphics[width=0.45\textwidth]{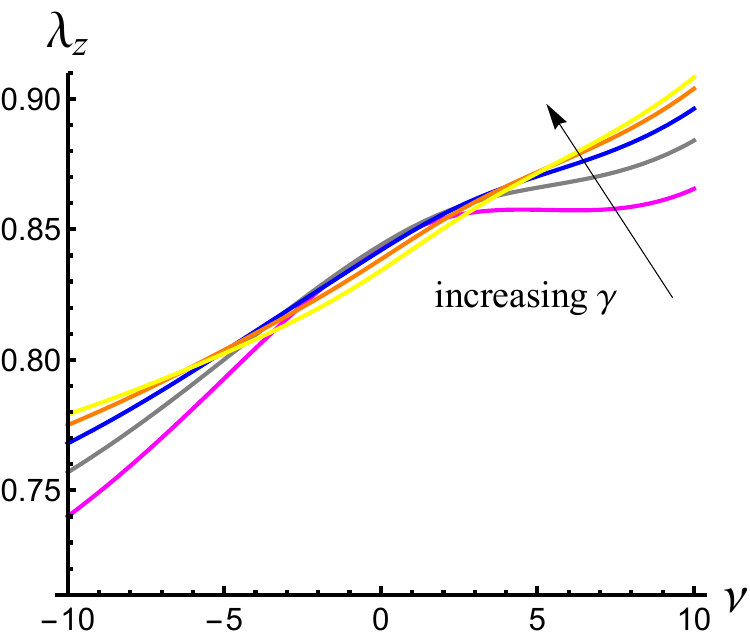}
    \caption{The critical stretch for mode $m = 2$ as a function of the dimensionless residual stress parameter $\nu$,  $\kappa = 0.5$ and $0.4 \leqslant\gamma\leqslant 0.6$  with increments of $\Delta \gamma=0.05$. The results shown on the left are for $B/A=1.2$, the ones on the right for $B/A=1.4$.}
    \label{fig:m=2-1}
\end{figure}

In Figure  \ref{fig:m=2-1} we report the critical stretch $\lambda_z$ for mode $m=2$, $\kappa=0.5$  as a function of the dimensionless residual stress parameter  $\nu$ for $0.4 \leqslant \gamma\leqslant 0.6$ with increments of $0.05$.  We again depict the results for $B/A=1.2$ in the left panel, for $B/A=1.4$ on the right. The curves on the left   show that the critical compression stretch  decreases almost linearly with $\nu$,  on the right  the effect is amplified. Therefore, the influence of residual stress increases with wall thickness.

\begin{figure}[!htbp]
    \centering
        \includegraphics[width=0.45\textwidth]{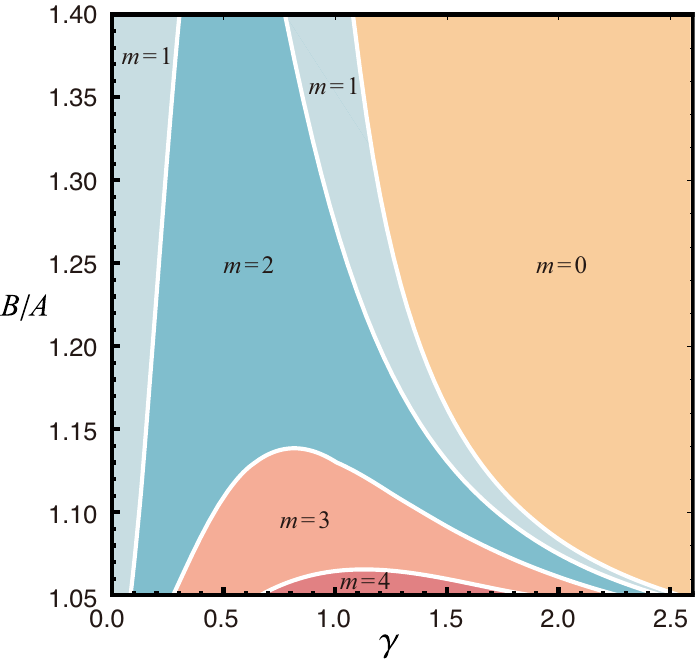}
        \includegraphics[width=0.45\textwidth]{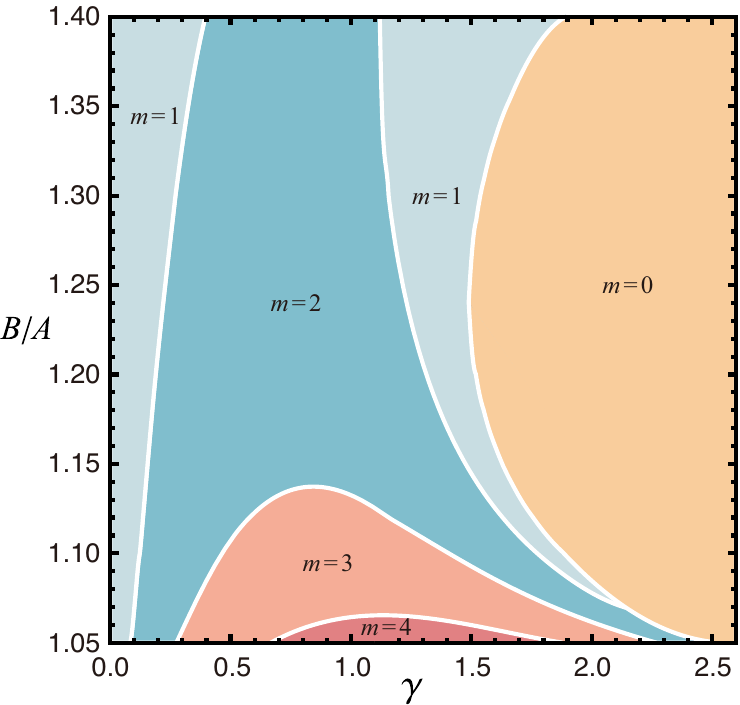}
         \caption{ Mode transitions  in the $B/A-\gamma$  plane when $\kappa=0.5$. The left image illustrates the response with no residual stress, the one on the right when the dimensionless residual stress parameter $\nu = 5$.}
            \label{fig:phase-diagram}
\end{figure}

Phase diagrams  are a convenient tool to illustrate mode transitions, which we develop next. \citet{goriely2008nonlinear} constructed a phase diagram for a tube with no residual stress  and captured the transitions from barreling  mode $m=0$, to  Euler buckling  $m=1$ and to mode $m=2$. Here, in the presence of a residual stress we take  $\kappa=0.5$ to reduce the number of free parameter to three. These are the slenderness ratio $\gamma$,  the thickness ratio $B/A$, and the dimensionless residual stress parameter $\nu$. In Figure 
\ref{fig:phase-diagram} we illustrate the transitions  in the $B/A-\gamma$  plane.  The left image illustrates the response with no residual stress, the one on the right when $\nu = 5$. We observe that for low values of $B/A$ higher order   modes appear and  that an increase in $\gamma$  amplifies the effect of the residual stress.

Consider, for example, a residually stressed tube with  $B/A=1.05$. Then, the image on the right in Figure \ref{fig:phase-diagram} shows that for small values of $\gamma$ the Euler buckling mode $m=1$ is energetically favorable. With increasing $\gamma$, higher order modes emerge and for short tubes with $\gamma > 2.5075$ the barreling mode $m=0$ dominates.  For a tube with $B/A = 1.3$  higher-order buckling modes are absent. We again find that the Euler buckling mode dominates for small values of  $\gamma$, while for short tubes  barreling  is preferred. 

\begin{figure}[htbp]
    \centering
    \subfigure[$m=0$]{
        \includegraphics[width=0.21\textwidth]{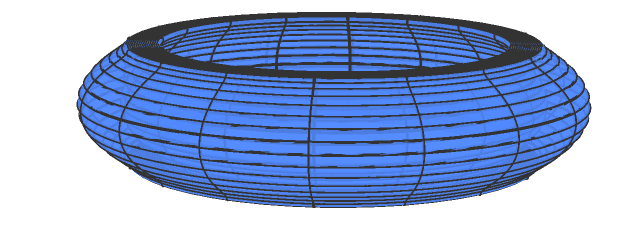}
        \label{fig:m=0,n=1}}
    \subfigure[$m=1$]{
        \includegraphics[width=0.14\textwidth]{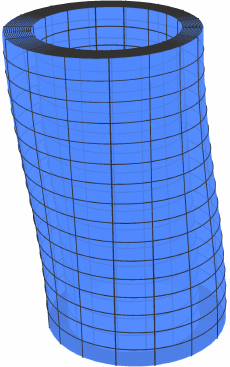}
        \label{fig:m=1,n=1}}
    \subfigure[$m=2$]{
        \includegraphics[width=0.18\textwidth]{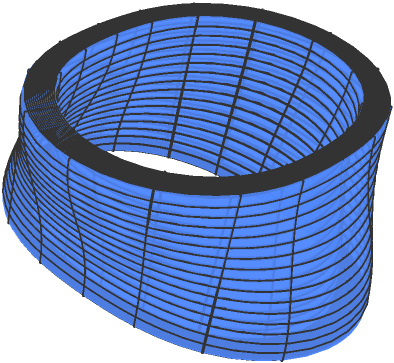}
        \label{fig:m=2,n=1}}
    \subfigure[$m=3$]{
        \includegraphics[width=0.18\textwidth]{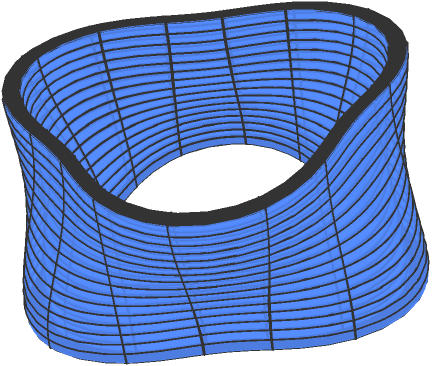}
        \label{fig:m=3,n=1}}
    \subfigure[$m=4$]{
        \includegraphics[width=0.18\textwidth]{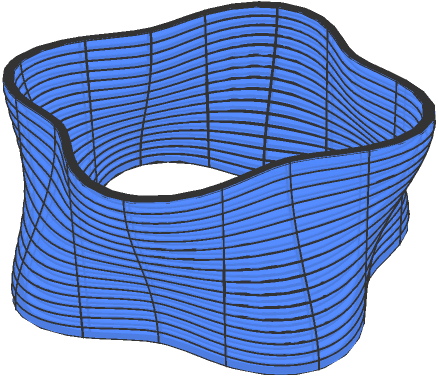}
        \label{fig:m=4,n=1}}
    \caption{Bifurcation modes  for different instabilities. Parameters are:   $\gamma$ = 2, $B/A$ = 1.2 in (a); $\gamma$ = 1, $B/A$ = 1.2 in (b); $\gamma$ = 1, $B/A$ = 1.1 in (c); $\gamma$ = 1, $B/A$ = 1.01 in  (d)  and  $\gamma$ = 0.3, $B/A$ = 1.35 in (e).}
    \label{fig:eigenmode}
\end{figure}

To provide further insight, we show the bifurcation modes of a compressed tube with no residual stress in Figure~\ref{fig:eigenmode}. It is well known that the linear buckling analysis does not provide the amplitude of the post-buckling states and,  for visualization purpose, an arbitrary  amplitude is assigned to each mode.

%%%%%%%%%%%%%%%%%%%%%%%%%%%%%%%%%%%%%%%%%%%%%%%%%%%%%%%%%%%%%%%%
\section{Conclusion}\label{conclusion}
%%%%%%%%%%%%%%%%%%%%%%%%%%%%%%%%%%%%%%%%%%%%%%%%%%%%%%%%%%%%%%%%

In this article we present a detailed stability analysis of a residually stressed tube under compression.  The governing equations and boundary conditions are formulated using the theory of nonlinear elasticity, following the works by \citep{hoger1985residual,  hoger1993constitutive, hoger1996elasticity}. To obtain explicit results, we specify the radial and circumferential residual stress components for a neo-Hookean material   \citep{sigaeva2019anisotropic, takamizawa2022stretch, ZHENG2016118}.

To conduct the buckling analysis, we superimpose an incremental displacement on the primary deformation and deduce an exact bifurcation condition given in Stroh form \citep{stroh1962steady,shuvalov2003frobenius,shuvalov2003sextic}. The buckling analysis is then reduced to studying the linearized increment in the deformation superimposed on a finitely deformed configuration subjected to a residual stress. To evaluate the effect of the residual stress we report the results of a  parametric study involving four  parameters. These are the wall thickness ratio $B/A$, the slenderness ratio $\gamma$,  the dimensionless residual stress  and  material parameters $\nu$ and $\kappa$. We determine the critical compression stretch $\lambda_z$ as a function of  the residual stress $\nu$, of the slenderness ratio $\gamma$ for  $B/A=1.2$ and $B/A=1.4$ for the stiffness parameter $\kappa=0.5$ and illustrate the results graphically. Three cases are considered: (i) the barreling mode $m=0$, the Euler buckling mode $m=1$ and the mode $m=2$. Higher order modes such as $m=3$ emerge in thin-walled tubes only and are therefore only briefly mentioned. In all  cases an increase in the wall-thickness amplifies the effect of the residual stress.

We also develop phase diagrams to illustrate  transitions from barreling $m=0$, to Euler buckling $m=1$ and to mode $m=2$. Responses with and without residual stress are compared and the results illustrated in the $B/A-\gamma$ plane. We show that for long tubes Euler buckling  is energetically favorable, for short tubes  the barreling mode dominates.

\section*{Acknowledgment}\label{Acknowledgment}
This work was supported by grants from the National Natural Science Foundation of China (Project Nos 12372072, 12072227 and 12021002). Y.L. further acknowledges the UKRI Horizon Europe Guarantee MSCA (Marie Sk\l odowska-Curie Actions) Postdoctoral Fellowship (EPSRC Grant No. EP/Y030559/1). For the purpose of Open Access, the author has applied a CC BY public copyright license to any Author Accepted Manuscript (AAM) version arising from this submission.

\bibliographystyle{elsarticle-harv}
\bibliography{references}

\end{document}